\documentclass[preprint,showpacs,preprintnumbers,amsmath,amssymb]{revtex4}

\usepackage{graphicx} 
\usepackage{bm}
\usepackage{epsfig}
\usepackage{color}
\usepackage{dcolumn} 

\newcommand{\ket}[1]{|#1\rangle}
\newcommand{\bra}[1]{\langle#1|}
\newcommand{\braket}[2]{\langle#1|#2\rangle}

\newcommand{\beq}{\begin{eqnarray*}}
\newcommand{\eeq}{\end{eqnarray*}}
\newcommand{\ie}{{\it i.e.}}
\newcommand{\eg}{{\it e.g.}}
\newcommand{\etc}{{\it etc.}}

\begin{document}

\setlength{\baselineskip}{0.70\baselineskip}


\title{The correlation energy functional within the $GW$-RPA approximation: exact forms, approximate forms and challenges}

\author{Sohrab Ismail-Beigi} \affiliation{Department of Applied Physics and Physics,
  Yale University, New Haven, CT 06520}
\date{\today}

\begin{abstract}
In principle, the Luttinger-Ward Green's function formalism allows one to compute simultaneously the total energy and the quasiparticle band structure of a many-body electronic system from first principles.  We present approximate and exact expressions for the correlation energy within the $GW$-RPA approximation that are more amenable to computation and allow for developing efficient approximations to the self-energy operator and correlation energy.  The exact form is a sum over differences between plasmon and interband energies.  The approximate forms are based on summing over screened interband transitions. We also demonstrate that blind extremization of such functionals leads to unphysical results: imposing physical constraints on the allowed solutions (Green's functions) is necessary.  Finally, we present some relevant numerical results for atomic systems.
\end{abstract}

\pacs{71.15.-m,71.15.Qe,71.15.Mb,71.15.Nc}
\keywords{Many-body Green's function,Luttinger-Ward,RPA,GW approximation,Ab Initio Calculations,Total energy,Self-energy}

\maketitle

\section{Introduction}
\label{sec:intro}

Density Functional Theory (DFT) \cite{HK,KS} with the local density (LDA) or generalized gradient approximations (GGA) \cite{KS,LDA,GGA} is the most widely used framework for first principles calculations of materials.  In practice, it is often found to provide a good description of ground-state total energies, atomic geometries, vibrational modes, {\it etc.} of a variety of materials.  A major shortcoming is its inability to predict accurate electronic band structure energies \cite{LDAbadgaps}.  For band insulators with weak correlation the failure is mainly quantitative \cite{HL}.  However, for the classic case of transition metal oxides, the failures can be qualitative and serious such as predicting a metallic instead of an insulating ground state \cite{LDApU}.  Ideally,  a computationally efficient {\it ab initio} method with accurate total energies {\it and} band structures would yield a major advance in predictive power.

A number of approaches aim to improve electronic band structures.  For systems with transition metals, one fundamental problem is that the LDA or GGA does not capture the proper electronic correlations for the spatially localized  $d$- and $f$-state derived bands.  Two current solutions to this deficiency are LDA+U \cite{LDApU} and dynamical mean-field theory (DMFT) \cite{DMFT1,DMFT2}.  The LDA+U approach is popular and easy-to-use: one ``manually'' adds static and localized correlation effects within a Hubbard-like model to the DFT energy functional.  More sophisticated is DMFT where localized but dynamic and high-level correlations are included using exact solutions of interacting quantum impurity models.  Both approaches are {\em physically} motivated in that they create frameworks that include to physics deemed important for the problem at hand.  A shared drawback is their requirement of an unspecified localized basis set in order to define key quantities such as the Hubbard parameters or the impurity site.  This raises questions of transferability, \ie, the dependence of their predictions on the chosen orbitals or parameter values.

A different approach is to use many-body perturbation theory of Green's functions.  The most successful is Hedin's $GW$ approximation to the electron self-energy \cite{Hedin}.  This approximation delivers high quality band structures of many band insulators and simple metals \cite{HL,TDDFTvsGWBSE}.  The $GW$ approximation includes a great deal of physics including exact exchange, localized Coulomb repulsion, dynamic screening, and dispersion forces: \eg, the LDA+U is a static approximation to $GW$ \cite{LDApU}.  However, most $GW$ calculations are perturbative in that they compute corrections to a mean-field DFT input, and the quality of the final result depends on the DFT description.  In certain transition metal oxides where the LDA provides a decent starting point, $GW$ corrections on this DFT starting point can yield a good description of the electronic bands \cite{GWNiO,TMOGWmodel,GWZnO,GWMgCaTiVO,GWaeSiMnONiO,GWNiO2}.  However, in other situations, the inadequacy of the DFT description can lead to  quantitative errors \cite{GWaeSiMnONiO,QPSCGW,scgwCu2O,scCOHSEX}.  

Clearly, it is advantageous to apply $GW$ beyond the perturbation-off-DFT prescription.  Such an approach would not assume a localized basis or any set of parameters.  Recent methods such as the Quasiparticle Self-Consistent $GW$ (QS$GW$) \cite{QPSCGW} or the self-consistent COHSEX (scCOHSEX) \cite{scCOHSEX} have successfully moved away from using DFT as the starting point.  These two methods find the noninteracting initial guess for the band structure approximately but self-consistently within $GW$.  Such approaches allow for inclusion of both static and dynamic screening effects in addition to some localized (Hubbard U) Coulombic physics in a single, general, and parameter-free framework.  QS$GW$ and scCOHSEX are self-consistent band structure methods, but it would be highly desirable to turn them into total energy methods via the Luttinger-Ward \cite{LW} approach that should, in principle, allow one to obtain accurate total energies and band structures.  

Separate from self-consistent band structure methods, there has been ongoing work on using $GW$-RPA type correlation functionals for computing total energies.  Much of the activity was sparked by initial work on the uniform electron gas showing that Luttinger-Ward functionals with the $GW$-RPA correlation energy provided very accurate ground-state total energies \cite{GWheg1,GWheg2,GWheg3,GWheg4}.  Other model calculations cast doubt on whether such high accuracy was generic \cite{GWhubbard}.  However, actual calculations on atoms, small molecules, and simple solids find ground-state energies that improve over the standard DFT functionals, especially when short-ranged corrections are added \cite{GWmolecules, GWH2hubbard,Miyake02,GWH2Be2, GWSiNa, DahlenvonBarthPRB04,DahlenLeeuwenvonBarth06,Marinivdw06,HellgrenvonBarth07,HarlKresse08,Toulouserangesep09}.  The vast majority of such calculations are post-DFT calculations.  Namely, they use Green's functions based on LDA or GGA wave functions and eigenvalues to evaluate the correlation energy instead of using a self-consistent one-particle description coming from a more elaborate, and presumably more accurate, theory like QS$GW$ or scCOHSEX.  In addition, the correlation energies are evaluated using the standard formula relying on frequency integration of the trace of a matrix logarithm (see Section~\ref{sec:wprewrite}).

In this work, we report on three main points of progress in this general area for the $GW$-RPA correlation energy functional.  The first is an exact rewriting of the correlation energy in terms of plasma and interband energies, a result that was found recently using very different methods \cite{Furche}.  This exact form is amenable to computations of identical complexity to present-day linear response time-dependent DFT (TDDFT) or Bethe-Salpeter (BSE) calculations \cite{TDDFT1,TDDFT2,TDDFT3,BSE,BSElong,TDDFTvsGWBSE}.  In addition, this new form has much better convergence properties when compared to the standard frequency integration method as evinced by our atomic calculations.  Second, we prove that the $GW$-RPA correlation energy functional is not bounded from below, has a minimum with negative infinite value, and when evaluated using non-interacting Green's functions,  it has no extrema.  This means blind optimization of total energy functionals that are based on the $GW$-RPA correlation energy is highly problematic, and physical constraints are required.  Third, we rewrite the $GW$-RPA correlation energy approximately as a sum over screened interband transition contributions.  This allows us to create a ladder of approximations of which the COHSEX \cite{Hedin} is the lowest wrung.  In addition, this allows us to put schemes such as scCOHSEX on a firm footing by showing how they can originate from a variational principle.  The ladder provides a series of more accurate functionals and associated self-energy operators that may deliver improved band structure and Green's functions.  We hope that these findings pave the way towards efficient and accurate computation of total energies within $GW$-RPA as well as the creation of efficient self-consistent schemes for computing total energies {\em and} band structures that improve over DFT.

This paper is organized as follows.  In Section \ref{sec:defs}, we describe our notation and provide basic definitions.  Section \ref{sec:lw} provides a brief description and the necessary ingredients of the Luttinger-Ward approach.  Section \ref{sec:noG0dep} describes how the total energy Klein functional we focus on in this work actually has no dependence on the choice of non-interacting auxiliary Green's function that enters the theory.  This greatly simplifies the form of the functional if we restrict ourselves to evaluating the total energy on non-interacting Green's functions.  We also make a connection to the Sham-Schl\"uter equation \cite{ShamSchluter,LSS} appropriate for the variational problem associated with the functional.  In Section \ref{sec:wprewrite}, we rewrite the $GW$-RPA correlation energy in terms of integrals of the standard, time-ordered RPA polarizability.  Section \ref{sec:derivs} provides expressions for the derivatives of the total energy functional versus the wave functions and eigenenergies or equivalently the static and Hermitian non-local potential determining the non-interacting Green's function.  In Section \ref{sec:exactrewritewp} we perform an exact rewriting of the $GW$-RPA correlation energy functional in terms of plasma and interband transition contributions.  Section \ref{sec:unbounded} provides a proof of the unboundedness and lack of extremum of the Klein total energy functional when evaluated on non-interacting Green's functions.  We discuss what this result means in more physical terms, how it relates to other results in the literature, and the requirement of constrained optimization that stems from this result.  
In Section \ref{sec:approxPhic}, we switch gears and derive approximate forms of the $GW$-RPA correlation energy written in terms of contributions from self-interaction-free screened interband transitions.  Section \ref{sec:cohsex} derives a ladder of approximations that allows us to connect to the COHSEX approximation for the self-energy and its associated correlation energy expression.  We provide a number of approximate correlation energies from which appropriate self-energies can be derived.  In Section \ref{sec:atoms}, we report numerical results for atoms demonstrating the lack of a lower bound to the correlation energy, the superior convergence properties of the exact plasmon form, and a tabulation of the quality of the various approximate forms derived in Sections \ref{sec:approxPhic} and \ref{sec:cohsex}.  We summarize and provide an outlook in Section \ref{sec:conclusions}.

\section{Definitions \& Notation}
\label{sec:defs}

In this work, we restrict ourselves to systems with time-independent, non-relativistic, many-body Hamiltonians so that all response or Green's functions are functions only of time differences.  Furthermore, we will consider systems with time reversal symmetry so that quantities such as Hamiltonians, density matrices, or one-particle states are real-valued.  We set $\hbar=1$ so energies and frequencies are interchangeable.

Wherever possible, we use matrix notation.  For example, the one-particle electron Green's function for a time-independent system in the frequency domain, $G(x,x',\omega)$, is a function of three arguments.  The $x$ and $x'$ arguments include both spatial coordinates and spin: $x=(\vec r, \sigma)$ where $\vec r$ is a three-vector and $\sigma=\pm 1$ labels the two spin projections.  In matrix notation, we write the matrix $G(\omega)$ whose matrix elements are $\bra{x}G(\omega)\ket{x'}=G(x,x',\omega)$.  

The time-ordered, non-interacting, one-particle Green's function $G_0(\omega)$ is given by
\begin{equation}
G_0(\omega) = (\omega I - H_0)^{-1} = \sum_n \frac{\ket{n}\bra{n}}{\omega-\epsilon_n}\,.
\label{G0def}
\end{equation}
The eigenenergies $\epsilon_n$ have imaginary parts  $Im\, \epsilon_n=\mbox{sgn}(\mu-\epsilon_n)0^+$ where $0^+$ is a positive infinitesimal and $\mu$ is the Fermi energy.  Thus the poles of $G_0$ are above the real $\omega$ axis for occupied or valence states, labeled by $v$ so $\epsilon_v<\mu$, and below the real axis for unoccupied or conduction states, labeled by $c$ so $\epsilon_c>\mu$.  The non-interacting one-particle Hamiltonian $H_0$ generates the orthonormal eigenstates $\ket{n}$ and real eigenvalues $\epsilon_n$,
\[
H_0 \ket{n} = \epsilon_n \ket{n}\,.
\]
The wave functions in coordinate space are denoted as $\psi_n(x)=\braket{x}{n}$.  In the time domain, we have the time-ordered formula
\begin{equation}
\mathcal{G}_0(t) = \int_{-\infty}^\infty \frac{d\omega}{2\pi} e^{-i\omega t}G_0(\omega) = -i\theta(t) \sum_c \ket{c}\bra{c}e^{-i\epsilon_c t} +i \theta(-t)\sum_v \ket{v}\bra{v} e^{-i\epsilon_vt}\,.
\label{G0t}
\end{equation}
The non-interacting density matrix is the standard sum over the occupied states
\[
\rho_0 = \sum_n \ket{n}\theta(\mu-\epsilon_n)\bra{n} = \sum_v \ket{v}\bra{v}
= -i\mathcal{G}_0(-0^+) = \int_{-\infty}^\infty \frac{d\omega}{2\pi i} e^{i\omega 0^+}G_0(\omega) \,.
\]
We separate a potential $U_0$ from $H_0$ via
\[
H_0 = T + U_{ion} + U_0
\]
where $T$ is the kinetic operator and $U_{ion}$ the electron-ion interaction potential.  The static and Hermitian $U_0$ represents approximately the effects of the Coulomb interaction  (Hartree, exchange, and correlation).  For example, in DFT $U_0$ is a local potential that is the sum of the Hartree and exchange-correlation potentials.  However, in general, we allow for a non-local $U_0$, \ie\ $U_0(x,x')\ne0$ for $\vec r\ne {\vec r}\,'$.  For a fixed nuclear configuration and thus $U_{ion}$, $G_0$ is determined by $U_0$ and vice versa:
\begin{equation}
G_0(\omega)^{-1} = \omega I - H_0 = \omega I - T - U_{ion} - U_0
\label{eq:G0U0relation}
\end{equation}

The exact, interacting, one-particle Green's function obeys the Dyson equation
\[
G(\omega)^{-1} = \omega I - T - U_{ion} - \phi_H - \Sigma_{xc}(\omega)
\]
where $\phi_H$ is the Hartree potential determined by the electron density $n(x)$
\[
\phi_H(x) =  \int dx'\ V(x,x')n(x')
\]
and the bare Coulomb operator is
\[
V(x,x') = \frac{\delta_{\sigma,\sigma'}}{|\vec r - \vec r\,'|}\,.
\]
The self-energy $\Sigma_{xc}(\omega)$ is frequency-dependent (dynamic) and non-Hermitian and encodes the complex exchange and correlation effects of the many-body system.  We can write the Dyson equation equally as
\begin{equation}
G^{-1}(\omega) = G_0^{-1}(\omega) - \Big[ \phi + \Sigma_{xc}(\omega) - U_0\Big]\,.
\label{eq:dysoneq}
\end{equation}
This shows that to obtain the true interacting Green's function, we replace the static, Hermitian $U_0$ by the dynamic, non-Hermitian $\phi_H+\Sigma_{xc}(\omega)$.    The interacting electron density $n(x)$ and density matrix $\rho(x,x')$ can be computed from the Green's function via
\begin{equation}
n(x) = \rho(x,x) \qquad , \qquad \rho(x,x') = \int_{-\infty}^\infty \frac{d\omega}{2\pi i} e^{i\omega 0^+} G(x,x',\omega) = -i\mathcal{G}(x,x',t= -0^+)
\label{eq:nrhoGrel}
\end{equation}
where the last form is the Green's function in the time domain evaluated for infinitesimal negative times.  Note that the relations of Eq.~(\ref{eq:nrhoGrel}) among $n$, $\rho$, $G$ and $\mathcal{G}$ also hold for the non-interacting $n_0$, $\rho_0$, $G_0$ and $\mathcal{G}_0$.

The frequency dependent dielectric matrix $\varepsilon(\omega)$ is related to the irreducible polarizability matrix $P(\omega)$ via 
\[
\varepsilon(\omega) = I - V P(\omega)\,.
\]
Within the RPA approximation, $P(\omega)$ is a sum over transitions between valence and conduction states,
\[
P(\omega) = \sum_{c,v} \frac{2(\epsilon_c-\epsilon_v)\ket{cv}\bra{cv}}{\omega^2-(\epsilon_c - \epsilon_v - i0^+)^2}\,.
\]
The pair states $\ket{cv}$ are defined in coordinate space via the single-particle wave functions $\psi_n(x)$ through
\[
\braket{x}{cv} \equiv \braket{x}{c}\braket{x}{v}^* = \psi_c(x)\psi_v(x)^*\,.
\]
For a more compact notation, we label each interband transition $(c,v)$ by $t$ with energy $\Delta_t \equiv \epsilon_c - \epsilon_v - i0^+$ and can write
\begin{equation}
P(\omega) = \sum_t \Pi_t(\omega) \qquad , \qquad \Pi_t(\omega) =  \frac{2\Delta_t\ket{t}\bra{t}}{\omega^2-\Delta_t^2}\,,
\label{eq:Ptdef}
\end{equation}
where $\Pi_t(\omega)$ is the polarizability contribution of transition $t$.  We can write $P(\omega)$ in a more general form to handle possible degeneracies in the transition energies $\Delta_t$.  Namely, we break the sum over all transitions $t$ into a first sum over the distinct energies $\Delta$ and then a sum over all degenerate transitions $\delta$ with energy $\Delta$:
\begin{equation}
P(\omega) = \sum_\Delta \Pi_\Delta(\omega) \ \ , \ \  \Pi_\Delta(\omega) =  \frac{2\Delta}{\omega^2-\Delta^2}\sum_{\delta\, |\, \Delta_\delta=\Delta} \ket{\delta}\bra{\delta}\,.
\label{eq:PDeltadef}
\end{equation}
Finally, the time domain RPA polarizability is
\begin{equation}
\mathcal{P}(x,x',t) = -i\mathcal{G}_0(x,x',t)\mathcal{G}_0(x',x,-t) = \int_{-\infty}^\infty \frac{d\omega}{2\pi} e^{-i\omega t}P(x,x',\omega)\,.
\label{Ptdef}
\end{equation}

\section{Luttinger-Ward approach}
\label{sec:lw}

One of the central points of the overall approach of Luttinger and Ward \cite{LW} is that one can obtain both the ground-state total energy and interacting $G(\omega)$ from the extremum of an energy functional of $G$.  In this work we concentrate on the specific case of the Klein functional \cite{Klein}, a functional of both $G$ and an auxiliary non-interacting $G_0$ that is meant to be an initial guess for $G$.  It is given by
\begin{multline*}
F[G,G_0] =\!\!\!\int_{-\infty}^\infty \frac{d\omega}{2\pi i} e^{i\omega 0^+}tr \Big\{ H_0G_0(\omega) +I  -G_0(\omega)^{-1}G(\omega)\\ 
+ \ln[G_0(\omega)^{-1}G(\omega)]-U_0G(\omega)  \Big\}
 + E_H[n]+ \Phi_{xc}[G]\,.
\label{eq:FLW}
\end{multline*}\
The $\omega$ integral can be turned into a closed contour integral over the upper complex $\omega$ plane due to the factor $e^{i\omega 0^+}$.  $E_H$ is the Hartree energy stemming from the electron density $n(x)$ associated with $G(\omega)$:
\[
E_H[n] = \frac{1}{2} \int dx \int dx'\ n(x)V(x,x')n(x') = \frac{1}{2} \int dx \ n(x) \phi_H(x)\,.
\]
The functional $\Phi_{xc}[G]$ is the exchange-correlation energy functional for this approach and is a complicated functional of $G$.  Formally, it is a sum over all diagrams to all orders in the Coulomb interaction obtained by closing all skeleton self-energy diagrams with Green's functions (with appropriate weight) \cite{LW}.  Much like in DFT, choosing an approximate form for $\Phi_{xc}$ corresponds to including a certain level of treatment of exchange-correlation effects.  However, since $\Phi_{xc}$ is a functional of the more information-rich, non-local and dynamic $G(x,x',\omega)$ as opposed to the simpler density $n(x)$ in DFT, relatively simple forms for $\Phi_{xc}$ will be equivalent to rather complex functionals of the density in DFT.  On the other hand, compared to DFT, the price for more information is the increased complexity of the energy functional $F$ and the entire theoretical and computational approach.

Within Luttinger-Ward theory, one focuses on the extremum of the functional $F$. At the extremum, the value of $F$ is the true ground-state total energy, and the extremizing $G$ is the true one-particle Green's function \cite{LW,Hedin}.  Therefore this framework provides, in principle, both exact total energies and one-particle properties such as quasiparticle wave functions, electron densities, and band structures.  To find the variation of $F$ versus $G$, we use a matrix differentiation rule based on the properties of the determinant:
\begin{equation}
tr \{ \ln(A) \} = \ln \{ \det(A) \}\qquad \rightarrow \qquad \delta \,tr \{ \ln(A) \} = 
tr \{ A^{-1}\delta A \}
\label{eq:logmatdef}
\end{equation}
where $\delta A$ is the variation in the matrix $A$.  The variation of $F$ for fixed $G_0$ is
\begin{eqnarray}
\delta F\Big|_{G_0} & = & \int_{-\infty}^\infty \frac{d\omega}{2\pi i} e^{i\omega 0^+} tr \Big\{ -G_0(\omega)^{-1}\delta G(\omega) + G(\omega)^{-1}\delta G(\omega) -U_0\,\delta G(\omega) \Big\} + \delta E_H + \delta \Phi_{xc}\nonumber\\
& = & \int_{-\infty}^\infty \frac{d\omega}{2\pi i} e^{i\omega 0^+}
tr \Big\{\left[
G(\omega)^{-1}-G_0(\omega)^{-1} + \phi_H + 2\pi i\frac{\delta \Phi_{xc}}{\delta G(\omega)} -U_0
\right] \delta G(\omega)
\Big\}\,.
\label{eq:dFdG}
\end{eqnarray}
Setting this to zero for arbitrary $\delta G$ yields the Dyson Eq.~(\ref{eq:dysoneq}) with the self-energy given by the functional derivative
\[
\Sigma_{xc}(\omega) = 2\pi i \frac{\delta \Phi_{xc}}{\delta G(\omega)}\,.
\]
Again, the situation is analogous to DFT where the exchange-correlation potential is the functional derivative versus the density of the exchange-correlation energy functional.

Within Luttinger-Ward theory, there are two separate challenges.  The first and most obvious is choosing some approximate form for $\Phi_{xc}$.  Second, there is the additional challenge that, unlike DFT where $N$-presentability conditions for the electron density $n(x)$ have been known for a long time \cite{Gilbert,Harriman}, similar conditions for the Green's function $G(x,x',\omega)$ are not known to the best of our knowledge.  In other words, it is not generally known which subset of functions $G(x,x',\omega)$ correspond to physically realizable Greens functions for interacting electrons with the standard many-body Hamiltonian.  Therefore, one must also decide on some scheme for restricting oneself to physically correct forms of $G$.  From a more pragmatic viewpoint, working with an arbitrary function $G(x,x',\omega)$ of three variables  is computationally very demanding so that any simplifying assumptions on the form of $G$ are enormously helpful in practice.  In what follows, we will restrict $G(\omega)$ to be of non-interacting form, and the next section explains why this is a sensible choice.

\section{Lack of dependence of $F$ on $G_0$}
\label{sec:noG0dep}

Clearly some approximations are required to make progress.  We will first replace the true interacting $G(\omega)$ by something simpler in order to reduce the search space for the extremum. Restricting to noninteracting Green's functions that are generated by Hermitian Hamiltonians is sensible from a physical viewpoint: we will try to search for the ``best'' noninteracting picture of the electronic system. As an added benefit, we can employ known algorithms for Hermitian and orthonormal eigensystems. 

In fact, this restriction greatly simplifies the structure of $F$.  This is because $F$ does not in fact depend on the choice of $U_0$ or equivalently $G_0$.  Namely, $\delta F/\delta U_0=0$ for fixed $G$.  Physically, this is sensible since the final result for $G$ should not depend on the arbitrary initial guess $G_0$.  For example, in the Dyson Eq.~(\ref{eq:dysoneq}), we remove $U_0$ and replace it by the self-energy.  To demonstrate this lack of dependence, we use Eq.~(\ref{eq:logmatdef}) to find the variation of $F$ versus $G_0$ at fixed $G$:
\begin{multline*}
\delta F\Big|_{G} = \int_{-\infty}^\infty \frac{d\omega}{2\pi i} e^{i\omega 0^+} tr \Big\{
[\delta H_0] G_0(\omega) + H_0 [\delta G_0(\omega)] - [\delta G_0(\omega)^{-1}]G(\omega)\\ + G(\omega)^{-1}G_0(\omega)[\delta G_0(\omega)^{-1}] G(\omega) - [\delta U_0] G(\omega)\,.
\Big\}
\end{multline*}
Using $G_0(\omega)=(\omega I-H_0)^{-1}$ and the variation of an inverse $\delta [A^{-1}] = -A^{-1} [\delta A] A^{-1}$,
we have
\begin{equation}
\delta H_0 = \delta U_0 \qquad , \qquad \delta G_0(\omega)^{-1} = -\delta U_0  \qquad , \qquad \delta G_0(\omega) = G_0(\omega)[\delta U_0]G_0(\omega)\,.
\label{eq:dH0dV0dG0}
\end{equation}
Plugging these in gives
\begin{multline*}
\delta F\Big|_{G} = \int_{-\infty}^\infty \frac{d\omega}{2\pi i} e^{i\omega 0^+} tr \Big\{
[\delta U_0] G_0(\omega) + H_0 G_0(\omega)[\delta U_0]G_0(\omega) + [\delta U_0]G(\omega)\\ - G(\omega)^{-1}G_0(\omega)[\delta U_0] G(\omega) - [\delta U_0] G(\omega)
\Big\}\,.
\end{multline*}
Using the cyclicity of the trace, this simplifies to
\[
\delta F\Big|_{G} = \int_{-\infty}^\infty \frac{d\omega}{2\pi i} e^{i\omega 0^+} tr \Big\{
G_0(\omega)H_0 G_0(\omega) \delta U_0
\Big\}  = \int_{-\infty}^\infty \frac{d\omega}{2\pi i} e^{i\omega 0^+}
\sum_n \frac{\epsilon_n\bra{n}\delta U_0\ket{n}}{(\omega-\epsilon_n)^2}
\]
where we have traced over the orthonormal basis of eigenstates $\ket{n}$ and used the diagonal nature of $H_0$ and $G_0$ in this basis.  The $e^{i\omega 0^+}$ factor has us close the integral over the upper $\omega$ complex half-plane,
\[
\delta F\Big|_{G} = \oint \frac{d\omega}{2\pi i}\sum_n \frac{\epsilon_n\bra{n}\delta U_0\ket{n}}{(\omega-\epsilon_n)^2}\,.
\]
The numerators have no $\omega$ dependence.  The standard contour integral for analytic $f(z)$
\begin{equation}
\oint \frac{dz}{2\pi i}\cdot \frac{f(z)}{(z-a)^k} = \frac{1}{(k-1)!}\frac{d^{k-1}f(z)}{dz^{k-1}}\Big|_{z=a}
\label{cauchy}
\end{equation}
applied to this case with $k=2$ gives a zero derivative and null result.  So we have our desired result
\begin{equation}
\frac{\delta F[G,G_0]}{\delta U_0}\Big|_{G} = 0\,.
\end{equation}

Hence, for a fixed $G$, we can evaluate $F$ at any valid $G_0$ without changing its value.  Physically, $F$ doesn't depend on the initial choice of non-interacting Hamiltonian.  Since we are restricting $G$ to be noninteracting, we may as well set $G=G_0$.  Then $F$ simplifies to
\begin{eqnarray}
F[G_0,G_0] & = &  \int_{-\infty}^\infty \frac{d\omega}{2\pi i} e^{i\omega 0^+} tr \Big\{ [T+U_{ion}]G_0(\omega) \Big\}
 + E_H[n_0]+ \Phi_{xc}[G_0]\nonumber\\
 & = & tr \Big\{ [T+U_{ion}] \rho_0 \Big\} + E_H[n_0]+ \Phi_{xc}[G_0]
\label{FG0G0}
\end{eqnarray}
We have the noninteracting kinetic, electron-ion, Hartree and exchange-correlation energies.  Operationally, the first three terms are identical to their counterparts in DFT.  Due to the extremal nature of $F$ about the extremizing $G$, $F[G_0,G_0]$ provides a variational estimate of the ground-state energy with the error being smallest for the ``best'' $G_0$. Except for $\Phi_{xc}$, the other energy terms depend only on the density matrix $\rho_0$.  Only $\Phi_{xc}$ depends on the energy eigenvalues $\epsilon_n$.

Before we end this section, we make brief comments on the functional derivative of $F[G_0,G_0]$ versus $G_0$ or equivalently versus the potential $U_0$.  Starting from Eq.~(\ref{eq:dFdG}) with $G=G_0$ or Eq.~(\ref{FG0G0}) and using the 
relation of $\delta G_0$ to $\delta U_0$ in Eq.~(\ref{eq:dH0dV0dG0}), we have at least three equivalent ways of writing the variation of $F[G_0,G_0]$:
\begin{eqnarray}
\delta F[G_0,G_0] & = & \int_{-\infty}^\infty \frac{d\omega}{2\pi i} e^{i\omega 0^+}
tr \Big\{\left[\phi_H + \Sigma_{xc}(\omega) -U_0\right] \delta G_0(\omega)
\Big\}\,.\label{eq:dFG0G0dG01}\\ 
& = & \int_{-\infty}^\infty \frac{d\omega}{2\pi i} e^{i\omega 0^+}
tr \Big\{\left[T+U_{ion}+\phi_H + \Sigma_{xc}(\omega)\right] \delta G_0(\omega)
\Big\}\,.\label{eq:dFG0G0dG02}\\
& = & \int_{-\infty}^\infty \frac{d\omega}{2\pi i} e^{i\omega 0^+}  tr \Big\{G_0(\omega)\left[\phi_H+\Sigma_{xc}(\omega)-U_0\right]G_0(\omega)\,\delta U_0\Big\}\,.
\label{eq:shamschluter}
\end{eqnarray}
Eq.~(\ref{eq:dFG0G0dG01}), written in terms of the variation $\delta G_0$, is used below when computing derivatives of $F$ versus the $\psi_n(x)$ and $\epsilon_n$ that characterize $G_0$.  Eq.~(\ref{eq:shamschluter}) is written in terms of the variation of the potential $U_0$ that generates $G_0$.  Setting $\delta F=0$ in Eq.~(\ref{eq:shamschluter}) for arbitrary $\delta U_0$ yields the matrix equation
\begin{equation}
 \int_{-\infty}^\infty \frac{d\omega}{2\pi i} e^{i\omega 0^+}  G_0(\omega)\left[\phi_H+\Sigma_{xc}(\omega)\right]G_0(\omega) = \int_{-\infty}^\infty \frac{d\omega}{2\pi i} e^{i\omega 0^+}  G_0(\omega)U_0G_0(\omega)
\label{eq:LSS}
\end{equation}
which is the linearized Sham-Schl\"uter (LSS) equation for the exchange-correlation potential operator $V_{xc}=U_0-\phi_H$ \cite{ShamSchluter,LSS}.  This is the condition determining the static and Hermitian $V_{xc}$ that most closely resembles the dynamic and non-Hermitian $\Sigma_{xc}(\omega)$. As Casida has noted, this is equivalent to saying that $V_{xc}$ is optimal in a variational sense  \cite{Casidavxcoptimal}. The LSS equation is most often discussed in the context of DFT where $V_{xc}(x)$ is a local function whereas what we have here is the more general case of a non-local Hermitian $V_{xc}(x,x')$.   Solving the LSS is nontrivial because (a) the frequency integral must be evaluated numerically, and (b) $G_0$ and $\Sigma_{xc}$ depend on $U_0$ in a nonlinear manner making for a self-consistent problem.   We will return to the question of local versus non-local $V_{xc}$ in Section~\ref{sec:unbounded} where we will see that solving this equation as written generates unphysical results, and we will discuss what is known in the literature for local $V_{xc}$ functionals.

\section{$\Phi_{xc}^{GW}$ in terms of the standard RPA polarizability}
\label{sec:wprewrite}

Beyond using an approximate or constrained form for $G$, we also require approximations for $\Phi_{xc}$.  In this work, we study $\Phi_{xc}$ within the $GW$-RPA approximation, normally defined via
\begin{equation}
\Phi^{GW}_{xc}[G_0] = 
\frac{1}{2}\int_{-\infty}^\infty \frac{d\omega}{2\pi i}\, tr \Big\{ \ln \bar\varepsilon(\omega) \Big\}\,.
\label{PhiGWxc}
\vspace*{-1ex}
\end{equation}
The dielectric matrix $\bar\varepsilon(\omega)$ is modified \cite{Hedin} and related to a modified polarizability $\bar P(\omega)$ via
\[
\bar\varepsilon(\omega) = I - V \bar P(\omega)\,.
\]
The modified $\bar P$ is most easily defined in the time domain
\[
\bar{\mathcal{P}}(t) = \int_{-\infty}^\infty \frac{d\omega}{2\pi} e^{-i\omega t}\bar P(\omega)
\]
as
\begin{equation}
\bar{\mathcal{P}}(x,x',t) = -i\mathcal{G}_0(x,x',t-0^+)\mathcal{G}_0(x',x,-t-0^+)\,.
\label{eq:Pbartdef}
\end{equation}
Specifically, since $\rho_0(x,x')=-i\mathcal{G}_0(x,x',t=-0^+)$, for zero time argument we have
\begin{equation}
\bar{\mathcal{P}}(x,x',0) = i\rho_0(x,x')\rho_0(x',x)\,.
\label{eq:Pbar0}
\end{equation}
This $\bar{\mathcal{P}}(t)$ is modified from the standard $\mathcal{P}(t)$ by the infinitesimal shifts in the arguments of $\mathcal{G}_0$ \cite{Hedin}: $\bar{\mathcal{P}}(t)$ and $\mathcal{P}(t)$ of Eq.~(\ref{Ptdef}) are identical for any finite non-zero $t$: they only differ in an infinitesimal neighborhood about $t=0$.  This ensures that the Fock exchange energy in $\Phi_{xc}^{GW}$ is properly recovered \cite{Hedin}.  
Hedin has provided formulae for $\Phi_{xc}$ including corrections beyond $GW$-RPA, so that from a formal point of view, it should be possible to proceed beyond the $GW$-RPA approximation \cite{Hedin}.

Computing $\Phi^{GW}_{xc}$ from Eq.~(\ref{PhiGWxc}) is cumbersome:  we must converge a continuous integral, and for each $\omega$ we need a matrix logarithm.  Furthermore, the physical meaning of the formula is hard to appreciate, making it difficult to create systematic approximations.   One of the main aims in this paper is to rewrite $\Phi^{GW}_{xc}$ in more tractable forms that permit us to understand its physical content.  

However, to proceed, we need to first rewrite Eq.~(\ref{PhiGWxc}) in terms of the usual time-ordered dielectric matrix $\varepsilon(\omega)=I-VP(\omega)$ where $P(\omega)$ is the standard RPA polarizability of Section \ref{sec:defs}.  We expand the logarithm in Eq.~(\ref{PhiGWxc}) to all orders using 
\[
\ln(1-x) = -\sum_{j=1}^\infty \frac{x^j}{j}
\]
to arrive at
\[
\Phi^{GW}_{xc} = 
-\frac{1}{2}\sum_{j=1}^\infty \int_{-\infty}^\infty \frac{d\omega}{2\pi i}\cdot\frac{1}{j}\, tr \Big\{ [V\bar P(\omega)]^j \Big\}\,.
\]
The $j=1$ term in the series is
\[
j=1 \mbox{  term  : }
-\frac{1}{2}\int_{-\infty}^\infty \frac{d\omega}{2\pi i} tr \Big\{ V\bar P(\omega)\Big\} =
\frac{i}{2}tr \Big\{ V \int_{-\infty}^\infty \frac{d\omega}{2\pi } \bar P(\omega)\Big\} = \frac{i}{2} tr \Big\{ V \bar{\mathcal{P}}(t=0) \Big\}\,.
\]
Writing the trace explicitly the $x$ basis and using Eq.~(\ref{eq:Pbar0}) shows that the $j=1$ term is just the Fock exchange energy $E_X$
\begin{equation}
j=1 \mbox{  term  : } -\frac{1}{2} \int dx \int dx' \ \rho_0(x,x')\rho_0(x',x)V(x,x') = E_X[\rho_0] \,.
\label{eq:EX}
\end{equation}
Thus, we have
\[
\Phi^{GW}_{xc} = E_X[\rho_0] 
-\frac{1}{2}\sum_{j=2}^\infty \frac{1}{j}\int_{-\infty}^\infty \frac{d\omega}{2\pi i}\, tr \Big\{ [V\bar P(\omega)]^j \Big\}\,.
\]
This has naturally separated $E_X$ from the correlation energy $\Phi_c$,
\[
\Phi_c = \Phi_{xc}-E_X\,.
\]
The remaining frequency integrals with $j\ge 2$ correspond to $j^{th}$ order autocorrelations of $\bar{\mathcal{P}}(t)$ evaluated at $t=0$, so the infinitesimal time shifts actually do not matter.  Specifically, the $j^{th}$ term has an $\omega$ integral 
\beq
\int_{-\infty}^\infty \frac{d\omega}{2\pi}\, tr \left\{ [V\bar P(\omega)]^j\right\} & = & \int_{-\infty}^\infty \frac{d\omega}{2\pi} \int_{-\infty}^\infty dt_1 \cdots \int_{-\infty}^\infty dt_j\, e^{i\omega(t_1+\cdots+t_j)} \, tr \left\{V\bar{\mathcal{P}}(t_1) \cdots V \bar{\mathcal{P}}(t_j)
\right\}\\
& = & \int_{-\infty}^\infty dt_1 \cdots \int_{-\infty}^\infty dt_j\, \delta(t_1+\cdots+t_j) \, tr \left\{V\bar{\mathcal{P}}(t_1) \cdots V \bar{\mathcal{P}}(t_j)
\right\}\,.
\eeq
We integrate over the times so that the infinitesimal region about $t=0$ where $\bar{\mathcal{P}}(t)\ne\mathcal{P}(t)$ is of zero measure.  Hence, we can safely replace $\bar{\mathcal{P}}$ by $\mathcal{P}$ for the $j\ge2$ terms  to get
\beq
\Phi^{GW}_{xc} & = & E_X[\rho_0] 
-\frac{1}{2}\sum_{j=2}^\infty \frac{1}{j}\int_{-\infty}^\infty \frac{d\omega}{2\pi i}\, tr \Big\{ [V P(\omega)]^j \Big\} \\
& = & E_X[\rho_0]  + \frac{1}{2}\int_{-\infty}^\infty \frac{d\omega}{2\pi i}\, tr \Big\{V P(\omega) \Big\} + \frac{1}{2}
\int_{-\infty}^\infty \frac{d\omega}{2\pi i}\, tr \Big\{ \ln\varepsilon(\omega) \Big\} \\
& = & E_X[\rho_0]  - \frac{i}{2} tr \Big\{V \mathcal{P}(t=0) \Big\} + \frac{1}{2}
\int_{-\infty}^\infty \frac{d\omega}{2\pi i}\, tr \Big\{ \ln\varepsilon(\omega) \Big\} \,.
\eeq
In the second line, we added and subtracted the corresponding $j=1$ term to the first line and summed up the series for the logarithm.  In the third line, evaluating $\mathcal{P}(t=0)$ requires some care as plugging in $t=0$ directly in Eq.~(\ref{Ptdef}) yields an indeterminate result.  Rather, we take the limit $t\rightarrow 0$ of $\mathcal{P}(t)$ in Eq.~(\ref{Ptdef}) to find
\[
\mathcal{P}(x,x',t\rightarrow 0) = -i\sum_c \psi_c(x)\psi_c(x')^* \sum_v \psi_v(x')\psi_v(x)^*
\]
or in matrix form
\[
\mathcal{P}(t\rightarrow 0) = -i\sum_{c,v} \ket{cv}\bra{cv} = -i\sum_t \ket{t}\bra{t}\,.
\]
The direction of approach of the limit, $t\rightarrow 0^+$ versus $t\rightarrow -0^+$, is immaterial as both yield the same answer for real valued Green's functions $G_0(\omega)$ due to time reversal symmetry.  We then have our desired result: $\Phi_{xc}^{GW}$ written in terms of the time-ordered $\varepsilon(\omega)$,
\begin{equation}
\Phi_{xc}^{GW}[G_0] = 
E_X[\rho_0]  - \frac{1}{2} \sum_t \bra{t}V\ket{t} + \frac{1}{2}
\int_{-\infty}^\infty \frac{d\omega}{2\pi i}\, tr \Big\{ \ln\varepsilon(\omega) \Big\} \,.
\label{Phixcepsilon}
\end{equation}
This expression is our main workhorse.  
In Sections \ref{sec:exactrewritewp} and \ref{sec:approxPhic}, we evaluate this integral exactly and approximately to generate alternative forms for $\Phi_{xc}^{GW}$.

\section{Derivatives of $F$ and $\Phi_{xc}^{GW}$}
\label{sec:derivs}

In this section, we provide expressions for the derivatives of the exchange-correlation functional $\Phi_{xc}^{GW}[G_0]$ and the total energy functional $F[G_0,G_0]$ versus the wave functions $\psi_n$ and eigenenergies $\epsilon_n$ or equivalently the potential $U_0$ that determine $G_0$.  The $\epsilon_n$ derivatives are used in Section~\ref{sec:unbounded} to prove the unboundedness of the energy functional.  Separately, these derivative expressions can prove useful to investigators contemplating variational approaches that extremize $F[G_0,G_0]$ which  require analytical derivatives. 

We begin with variations of the eigenenergies $\epsilon_n$.  The derivative of $G_0$ is
\[
\frac{\partial G_0(\omega)}{\partial \epsilon_n} =  \frac{\ket{n}\bra{n}}{(\omega-\epsilon_n)^2}\,.
\]
As per Section \ref{sec:noG0dep}, the only non-zero contribution to the variation of $F[G_0,G_0]$ when changing $\epsilon_n$ comes from the exchange-correlation functional $\Phi_{xc}^{GW}$ so
\[
\frac{\partial F[G_0,G_0]}{\partial \epsilon_n} = \frac{\partial \Phi_{xc}^{GW}[G_0]}{\partial \epsilon_n} = 
\int_{-\infty}^\infty \frac{d\omega}{2\pi i} e^{i\omega 0^+} tr \Big\{
\Sigma_{xc}(\omega)\frac{\partial G_0(\omega)}{\partial \epsilon_n}\Big\}\,.
\]
We turn this into a contour integral over the upper complex $\omega$ half plane.
Using Eq.~(\ref{cauchy}), we get contributions from the poles of $\Sigma_{xc}(\omega)$ that are above the real axis and a possible contribution from the pole of $\partial G_0/\partial\epsilon_n$ if $n$ is occupied (if $\epsilon_n<\mu$).

To organize the process, we write $\Sigma_{xc}(\omega)$ in the general form of a sum over poles 
\begin{equation}
\Sigma_{xc}(\omega) = \Sigma_x + \sum_\alpha \frac{\sigma^+_\alpha}{\omega-\xi^+_\alpha} + \sum_\beta \frac{\sigma^-_\beta}{\omega-\xi^-_\alpha}\,.
\label{eq:sigmapoleform}
\end{equation}
Here, $\Sigma_x$ is the Fock (bare) exchange operator
\[
\Sigma_x(x,x') = -\sum_v\psi_v(x)\psi_v(x')^*V(x,x')
\]
The $\xi^+_\alpha$ are the poles of $\Sigma_{xc}(\omega)$ that are above the real axis, $Im\ \xi^+_\alpha>0$, and $\sigma_\alpha^+$ are their residues, while $\sigma^-_\beta$ and $\xi^-_\beta$ are the residues and poles below the real axis, $Im\ \xi^-_\beta<0$.  This allows us to write the derivative as
\[
\frac{\partial F[G_0,G_0]}{\partial \epsilon_n} = \sum_\alpha
\frac{\bra{n}\sigma^+_\alpha\ket{n}}{(\xi^+_\alpha - \epsilon_n)^2} +
\theta(\mu-\epsilon_n) \bra{n}\frac{d\Sigma_{xc}(\omega)}{d \omega}\ket{n}\Big|_{\omega=\epsilon_n}\,.
\]
A quick manipulation yields
\begin{equation}
\frac{\partial F[G_0,G_0]}{\partial \epsilon_n} = \theta(\epsilon_n-\mu)\sum_\alpha
\frac{\bra{n}\sigma^+_\alpha\ket{n}}{(\xi^+_\alpha - \epsilon_n)^2} -
\theta(\mu-\epsilon_n) 
\sum_\beta\frac{\bra{n}\sigma^-_\beta\ket{n}}{(\xi^-_\beta - \epsilon_n)^2}\,.
\end{equation}
If the residue matrices $\sigma^\pm$ are positive definite, then the derivatives for empty states are always positive while those of filled states are always negative.  For the $GW$-RPA approximation, we can demonstrate this explicitly: as we will discuss in more detail in in Section~\ref{sec:exactrewritewp}, the RPA screened Coulomb interaction $W(\omega)=\varepsilon^{-1}(\omega)$ has the outer product form
\[
W^{RPA}(\omega) = V + V\sum_p \frac{2\omega_p \ket{p}\bra{p}}{\omega^2-\omega_p^2}V\,.
\]
A standard exercise in time-dependent perturbation theory shows that the exact many-body $W$ also has the same outer product form
\[
W^{exact}(\omega) = V + V\sum_s \frac{2(E_s-E_0)\ket{n_{s0}}\bra{n_{s0}}}
{\omega^2-(E_s-E_0)^2}V\,.
\]
where $E_0$ is the exact ground-state energy, $E_s$ are the exact excited-state energies, $\braket{x}{n_{s0}}=\bra{0}\hat n(x)\ket{s}$ where $\hat n(x)$ is the fermion density operator and $\ket{0}$ and $\ket{s}$ are the many-body eigenstates.  Thus the exact and  RPA $W$ are related via the replacements $p,\omega_p,\ket{p}\rightarrow s,E_s-E_0,\ket{n_{s0}}$.

Whether we have the exact or RPA $W$, the $GW$ self-energy is 
\[
\Sigma_{xc}(x,x',\omega) = i \int_{-\infty}^\infty \frac{d\omega'}{2\pi}e^{-i\omega'0^+}G(x,x',\omega-\omega')W(x,x',\omega')\,.
\]
For $G=G_0$ and the RPA $W$, the integral yields
\[
\Sigma_{xc}(x,x',\omega) = \Sigma_x(x,x') + \sum_{p}
\bra{x}V\ket{p}\bra{p}V\ket{x'}\left(
\sum_v\frac{\psi_v(x)\psi_v(x')^*}{\omega+\omega_p-\epsilon_v}
+
\sum_c\frac{\psi_c(x)\psi_c(x')^*}{\omega-\omega_p-\epsilon_c}
\right)\,.
\]
Thus the poles and residues of the $GW$ $\Sigma_{xc}$ are indexed by an eigenvalue $\epsilon_n$ and a plasmon $\omega_p$: the labels $\alpha$ or $\beta$ in Eq.~(\ref{eq:sigmapoleform}) correspond  a pair index $(n,p)$.  The precise correspondence for the $GW$-RPA approximation is
\begin{equation}
\sigma^+_{v,p}(x,x') = \bra{x}V\ket{p}\bra{p}V\ket{x'}\psi_v(x)\psi_v(x')^* \ \ \ \ , \ \ \ \xi^+_{v,p} = \epsilon_v-\omega_p
\label{eq:corresp1}
\end{equation}
and
\begin{equation}
\sigma^-_{c,p}(x,x') = \bra{x}V\ket{p}\bra{p}V\ket{x'}\psi_c(x)\psi_c(x')^* \ \ \ \ , \ \ \ \xi^-_{c,p} = \epsilon_c+\omega_p\,.
\label{eq:corresp2}
\end{equation}
We arrive at the desired derivative
\begin{equation}
\frac{\partial F^{GW}[G_0,G_0]}{\partial \epsilon_n} = \theta(\epsilon_n-\mu)\sum_{v,p}
\frac{|\bra{nv}V\ket{p}|^2}{(\epsilon_n - \epsilon_v +\omega_p)^2} -
\theta(\mu-\epsilon_n) 
\sum_{c,p}\frac{|\bra{nc}V\ket{p}|^2}{(\epsilon_c-\epsilon_n+\omega_p)^2}\,.
\label{eq:dFdepsilonn}
\end{equation}
It is clear that for unoccupied states the derivative is always positive and for filled states it is always negative.  Repeating the derivation with the exact $W$ produces the same conclusion which is thus inherent to the $GW$ form of the self-energy when $G$ is non-interacting.  We discuss the implications of this result in Section~\ref{sec:unbounded}.

For completeness, we provide the derivatives of $F$ versus the wave functions $\psi_n(x)$ and versus the potential $U_0$.  For the wave function derivatives we have
\[
\frac{\delta F[G_0,G_0]}{\delta \psi_n(x)^*} = 
\int_{-\infty}^\infty \frac{d\omega}{2\pi i} e^{i\omega 0^+} tr \Big\{
[T+U_{ion}+\phi_H+\Sigma_{xc}(\omega)]\frac{\delta G_0(\omega)}{\delta \psi_n(x)^*}\Big\}\,.
\]
The functional derivative of $G_0$ is 
\[
\frac{\delta G_0(x',x'',\omega)}{\delta \psi_n(x)^*} = \frac{\psi_n(x')\delta(x''-x)}{\omega-\epsilon_n-i0^+\mbox{sgn}(\mu-\epsilon_n)}\,.
\]
We insert this into the integral and perform the contour integral.   For the general case, 
\begin{eqnarray}
\frac{\delta F[G_0,G_0]}{\delta \psi_n(x)^*} & = & 
\theta(\mu-\epsilon_n) \bra{x}T+U_{ion}+\phi_H+\Sigma_{xc}(\epsilon_n)\ket{n} + \sum_\alpha \frac{\bra{x}\sigma^+_\alpha\ket{n}}{\xi^+_\alpha-\epsilon_n}\\
& = & \theta(\mu-\epsilon_n) \bra{x}T+U_{ion}+\phi_H+\Sigma_{x}\ket{n}
\nonumber \\
& & +\ \theta(\epsilon_n-\mu)\sum_\alpha\frac{\bra{x}\sigma^+_\alpha\ket{n}}{\xi^+_\alpha-\epsilon_n}
-\theta(\mu-\epsilon_n)\sum_\alpha\frac{\bra{x}\sigma^-_\alpha\ket{n}}{\xi^-_\alpha-\epsilon_n}\,,
\end{eqnarray}
and expressions for the $GW$-RPA approximation are found by performing the substitutions of Eqs.~(\ref{eq:corresp1},\ref{eq:corresp2}).  
For the derivative versus $U_0$, it is convenient to express the potential in the one-particle eigenbasis so ${U_0}_{j,k}=\bra{j}U_0\ket{k}$.  The variation $\delta U_0$ is then
\[
\delta U_0 = \sum_{j,k} \ket{j}\, \delta {U_0}_{j,k}\, \bra{k}\,.
\]
Inserting this into Eq.~(\ref{eq:shamschluter}) and performing the contour integral and some algebraic manipulations yields
\begin{eqnarray}\frac{\partial F[G_0,G_0]}{\partial {U_0}_{v,v'}} & = & - \sum_\alpha \frac{\bra{v'}\sigma_\alpha^-\ket{v}}{(\xi_\alpha^--\epsilon_{v'})(\xi_\alpha^--\epsilon_v)}\\
\frac{\partial F[G_0,G_0]}{\partial {U_0}_{c,v}} & = & \frac{\bra{v}\phi_H+\Sigma_{xc}(\epsilon_v)-U_0\ket{c}}{\epsilon_v-\epsilon_c} + \sum_\alpha \frac{\bra{v}\sigma_\alpha^+\ket{c}}{(\xi_\alpha^+-\epsilon_{v})(\xi_\alpha^+-\epsilon_c)}\\
\frac{\partial F[G_0,G_0]}{\partial {U_0}_{v,c}} & = & 
\frac{\bra{c}\phi_H+\Sigma_{xc}(\epsilon_v)-U_0\ket{v}}{\epsilon_v-\epsilon_c}
+ \sum_\alpha \frac{\bra{c}\sigma_\alpha^+\ket{v}}{(\xi_\alpha^+-\epsilon_{c})(\xi_\alpha^+-\epsilon_v)}\\
\frac{\partial F[G_0,G_0]}{\partial {U_0}_{c,c'}} & = & \sum_\alpha \frac{\bra{c'}\sigma_\alpha^+\ket{c}}{(\xi_\alpha^+-\epsilon_{c'})(\xi_\alpha^+-\epsilon_c)}\,.
\end{eqnarray}
where $v,v'$ label valence states and $c,c'$ label conduction states.  As a check, the diagonal cases $v=v'$ or $c=c'$ yield the same result as the $\epsilon_n$ derivative as necessary by first order perturbation theory.  Setting all the derivatives to zero is equivalent to solving the LSS Eq.~(\ref{eq:shamschluter}) with no constraints on $U_0$.  This turns out to yield unphysical results as explained in Section~\ref{sec:unbounded}.

\section{$\Phi_{xc}^{GW}$ rewritten exactly with plasmon and interband energies}
\label{sec:exactrewritewp}

In this section, we rewrite $\Phi_{xc}^{GW}$ exactly in terms of a sum over plasmon and interband transition energies.  This exact expression has been reported by other workers using different methods \cite{Furche}.  (We found this result contemporaneously and independently.)  Our derivation is based on contour methods which we present for completeness before moving on to the implications of this exact expression.

We start with the result of Eq.~(\ref{Phixcepsilon}).  The time-ordered RPA $\varepsilon(\omega)$ is given by
\[
\varepsilon(\omega) = I - VP(\omega) = I - V \sum_t \frac{2\Delta_t\ket{t}\bra{t}}{\omega^2-\Delta_t^2}
\]
which has poles at the interband energies $\omega=\Delta_t$.  The poles of the inverse dielectric function $\varepsilon(\omega)^{-1}$ occur at the plasma frequencies which are the natural modes of collective charge oscillation.   The matrix $\varepsilon(\omega)^{-1}$ is given by
\begin{equation}
\varepsilon(\omega)^{-1} = I + V\chi(\omega) = I + V \sum_p \frac{2\omega_p \ket{p}\bra{p}}{\omega^2-\omega_p^2}\,.
\label{epsinvedef}
\end{equation}
(The $\omega_p$ have infinitesimal negative imaginary parts.) 
The plasma frequencies $\omega_p$ and mode functions $\ket{p}$ obey the RPA/Casida Hermitian eigenvalue equations \cite{Casida}
\begin{equation}
\Delta_t^2 C_{t,p} + \sum_t 2\sqrt{\Delta_t\Delta_{t'}}\bra{t}V\ket{t'}C_{t',p} = \omega_p^2 C_{t,p}\,.
\label{RPACasida}
\end{equation}
The column vectors $C_{t,p}$ are orthonormal and the $\ket{p}$ are given by
\[
\ket{p} = \sum_t \ket{t}\sqrt{\frac{\Delta_t}{\omega_p}} C_{tp}\,.
\]
We see that the $\omega_p$ are the eigenvalues of the square root of the $\Omega^2$ matrix defined as
\begin{equation}
\Omega^2_{t,t'} = \Delta_t^2 \delta_{t,t'} + 2\sqrt{\Delta_t\Delta_{t'}}\bra{t}V\ket{t'}\,.
\label{eq:Omega2matrix}
\end{equation}
Before going forward, we make a clarification concerning our use of the terms ``plasmon'' or ``plasma'' modes or frequencies.  In this work, these terms refer to solutions of the Casida/RPA Eq.~(\ref{RPACasida}) so that the $\omega_p$ are real frequencies.  Of course, a plasmon excitation in an actual material has finite life time due to damping processes.   Thus, we are dealing with the real part of the plasma frequencies, and this approximation is analogous to our assumption of a non-interacting Green's function where the quasiparticles have real eigenenergies and infinite lifetimes.

Since the poles of $\varepsilon(\omega)^{-1}$ are at $\omega_p$, its inverse $\varepsilon(\omega)$ has a zero at the $\omega_p$, \ie\ $\varepsilon(\omega_p)$ has a zero eigenvalue and is not invertible.  Therefore, for $\omega=\Delta_t$ one of the eigenvalues of $\varepsilon(\omega)$ diverges while for $\omega=\omega_p$ one of its eigenvalues is zero.  The number of transitions $t$ and plasmons $p$ are equal because Eq.~(\ref{RPACasida}) is a square matrix Hermitian eigenproblem.

Letting the eigenvalues of $\varepsilon(\omega)$ be $\lambda_m(\omega)$, the trace over the logarithm of $\varepsilon(\omega)$ is 
\[
\Phi_{xc}^{GW}[G_0] = 
E_X[\rho_0]  - \frac{1}{2} \sum_t \bra{t}V\ket{t} + \frac{1}{2}
\int_{-\infty}^\infty \frac{d\omega}{2\pi i}\sum_m \ln \lambda_m(\omega)\,.
\]
We close the integral on the complex $\omega$ lower half plane to yield a closed contour integral.  The function $\ln z$ is analytic everywhere except on the branch cut starting at $z=0$ and extending to $z=-\infty$ along the negative real axis.  This means that the integrand is analytic everywhere except between the zeros and poles of $\lambda_m(\omega)$, \ie\ between some interband energy $\Delta_t$ where 
\[
\lambda_m(\omega\rightarrow\Delta_t) = \frac{\mbox{constant}}{\omega-\Delta_t} + O((\omega-\Delta_t)^0)
\]
and some plasma frequency $\omega_p$ where
\[
\lambda_m(\omega\rightarrow\omega_p) = \mbox{constant}\cdot(\omega-\omega_p) + O((\omega-\omega_p)^2)\,.
\]
Due to the positive definite nature of the Coulomb interaction matrix $\bra{t}V\ket{t'}$ in Eq.~(\ref{RPACasida}), the smallest $\Delta_t$ is smaller than the smallest $\omega_p$ while the largest $\omega_p$ is larger than the largest $\Delta_t$.  Hence, we have a set of overlapping branch cuts beginning at the $\Delta_t$ and terminating at the $\omega_p$.  The contour integral collapses to an integral around the finite segments of length $\omega_p-\Delta_t$.  We perform the contour integral by remembering that $\ln z$ changes its imaginary part by $2\pi i$ when going from above to below the branch cut.  Each integral about a segment yields precisely $\omega_p-\Delta_t$. This leads to our first central result which is an exact rewriting of $\Phi_{xc}^{GW}$ in terms of plasma and interband energies:
\begin{equation}
\Phi_{xc}^{GW}[G_0] = 
E_X[\rho_0]  + \frac{1}{2}
\sum_p \omega_p - \frac{1}{2} \sum_t \Big(\Delta_t +\bra{t}V\ket{t}\Big) \,.
\label{wprewrite}
\end{equation}
Since we sum over all $t$ and $p$, the precise pairings of $\Delta_t$ and $\omega_p$ for each branch cut are irrelevant for the final result.

This exact form is much in the spirit of early work on the electron gas showing that the RPA correlation energy represents replacing interband oscillators by plasma modes \cite{Pines}. We refer the interested reader to Ref.~\cite{Furche} for a more detailed discussion of this and related points.  From a pragmatic viewpoint, this expression is amenable to direct computation since the RPA/Casida or related equations are solved regularly when performing TDDFT or BSE calculations \cite{BSElong,TDDFTvsGWBSE}.  Direct diagonalization of the RPA/Casida Eq.~(\ref{RPACasida}) makes for $O(N^6)$ scaling for a system of $N$ atoms.  However, since only the trace over all modes is needed, there are possibilities for improved scaling.  For example, the sum over plasma frequencies is also the trace of the square root of the $\Omega^2$ matrix,
\begin{equation}
\sum_p \omega_p = tr \left\{ \left(\Omega^2\right)^{1/2} \right\}\,,
\end{equation}
so efficient matrix square roots algorithms that avoid diagonalization are computationally advantageous \cite{Furche}.

In terms of convergence, Eq.~(\ref{wprewrite}) is superior to the standard integral form of Eqs.~(\ref{PhiGWxc}) or (\ref{Phixcepsilon}).  There is only a single convergence parameter which is the size of the basis set of transitions $\{\ket{t}\}$ in the calculation.  For the integral forms, one must converge additionally the continuous $\omega$ integral by dealing with issues of grid spacings, truncation at large $\omega$, \etc\  In Section~\ref{sec:atoms}, we will highlight the superior convergence properties of Eq.~(\ref{wprewrite}) numerically for atoms.

\section{Unboundedness of $\Phi_{xc}^{GW}$ and lack of extremum of $F$}
\label{sec:unbounded}

In this section, we prove that $\Phi_{xc}^{GW}[G_0]$ is in fact unbounded from below:  its minimum value is negative infinity.  Furthermore, the total energy functional $F$ can not have an extremum when optimized freely over non-interacting Green's functions $G_0$ that are generated by non-local potentials $U_0$.  We then discuss the meaning and consequences of these facts.

Beginning with the total energy functional of Eq.~(\ref{FG0G0}), we see that the only term depending on the $\epsilon_n$ is the exchange-correlation part $\Phi_{xc}[G_0]$.  For the $GW$-RPA approximation, we have the exact expression of Eq.~(\ref{wprewrite}) for $\Phi^{GW}_{xc}$.  This functional is unbounded from below if varied freely over $\epsilon_n$.  Consider sending all the eigenvalues towards the Fermi energy, $\epsilon_n\rightarrow\mu$.  This makes all transition energies tend to zero from above, $\Delta_t\rightarrow0^+$. The RPA/Casida Eq.~(\ref{RPACasida}) immediately shows that the plasmon energies must also tend to zero $\omega_p\rightarrow0^+$ as well.  Thus $\Phi_{xc}^{GW}$ will turn into
\[
\Phi_{xc}^{GW}[G_0] \rightarrow 
E_X[\rho_0]  - \frac{1}{2} \sum_t \bra{t}V\ket{t} = 
E_X[\rho_0]  - \frac{1}{2} \sum_{v}\sum_c \bra{cv}V\ket{cv} \,.
\]
Writing this out explicitly in coordinate space
\[
\Phi_{xc}^{GW}[G_0] \rightarrow 
E_X[\rho_0]  - \frac{1}{2} \sum_{v}\sum_c \int dx \int dx'\ 
\psi_c(x)^*\psi_v(x) V(x,x')\psi_c(x')\psi_v(x')^* \,.
\]
Using completeness
\[
\sum_c \psi_c(x)\psi_c(x')^* = \delta(x-x') - \sum_{v'}
\psi_{v'}(x)\psi_{v'}(x')^*
\]
where $v'$ labels occupied states, we find
\[
\Phi_{xc}^{GW}[G_0] \rightarrow 
E_X[\rho_0]  - \frac{1}{2} \sum_{v} \int dx\ 
\psi_v(x)V(x,x)\psi_v(x)
+\frac{1}{2}\sum_{v,v'}\bra{vv'}V\ket{vv'}\,.
\]
Since $V(x,x')=\delta_{\sigma,\sigma'}/|r-r'|$ diverges to positive infinity for $x=x'$, we conclude that $\Phi_{xc}^{GW}$ diverges to negative infinity when all eigenvalues approach the Fermi energy.  This particular choice of variation of the $\epsilon_n$ is perfectly permissible if one allows for arbitrary variations of the non-local potential $U_0$.  We give numerical evidence of this behavior for atoms in Section~\ref{sec:atoms}.

Therefore, global minimization of $F[G_0,G_0]$ over all allowed $G_0$ --- more precisely over all non-local potentials $U_0$ that generate the $G_0$ --- is guaranteed to give unphysical results.  However, one may still hope to find a local minimum or an extremum:  after all, the Luttinger-Ward approach only guarantees that the true $G$ is an extremum of the energy functional.  However, this is also not possible because $\partial F[G_0,G_0]/\partial \epsilon_n$ is never zero.  Referring back to Eq.~(\ref{eq:dFdepsilonn}) in Section~\ref{sec:derivs}, we see that the derivative versus $\epsilon_n$ is always positive for unoccupied states and always negative for occupied states.  Hence, a search for an extremum also drives us to the unphysical global minimum. No other extremum exists because the derivatives are never zero.  This proof confirms the suggestion \cite{QPSCGWneginf} that direct minimization of the $GW$-RPA total energy will give meaningless results.  

We now discuss the meaning and relation of this result to what is known in the literature.  A number of studies have used local (in coordinate $x$) exchange-correlation potentials $V_{xc}=U_0-\phi_H$ when generating $G_0$ for atoms and molecules.  The choices have included LDA, GGA, or optimized effective potentials  \cite{DahlenvonBarthPRB04,DahlenvonBarthJCP04,DahlenLeeuwenvonBarth06}.  In addition, a more recent work \cite{HellgrenvonBarth07} has solved the linearized Sham-Sch\"ulter (LSS) equation for spherical atoms to find the optimal local $V_{xc}$.  No sign of any instability has been found in any of these works and the numerical evidence shows that the LSS-optimal local $V_{xc}$ locates an energy minimum \cite{HellgrenvonBarth07}.  In addition, there are strong theoretical arguments for why the total energy functional will have a minimum when varied over local potentials \cite{DahlenLeeuwenvonBarth06}.

Unfortunately, the optimal local potential is not necessarily the best potential in terms of physical predictions.  In practice, using a non-local potential to generate $G_0$ can greatly improve results:  \eg, for atoms and diatomic molecules, the Hartree-Fock non-local potential yields total energies that are essentially indistinguishable from those given by using the exact, self-consistent Green's function $G$ \cite{DahlenLeeuwenvonBarth06}.  Therefore, including some nonlocality in the potential $U_0$ is important physically:  for atoms and small molecules, the Hartree-Fock description is significantly closer to the true Green's function than any description based on a local potential \cite{DahlenLeeuwenvonBarth06}.  This result is sensible since the true Green's function $G$ obeying Dyson's Eq.~(\ref{eq:dysoneq}) is generated by a non-local self-energy operator $\Sigma_{xc}(\omega)$, and we would expect a non-local $U_0$ to be a better approximation than a local operator.

\begin{figure}[t!]
\includegraphics[width=3in]{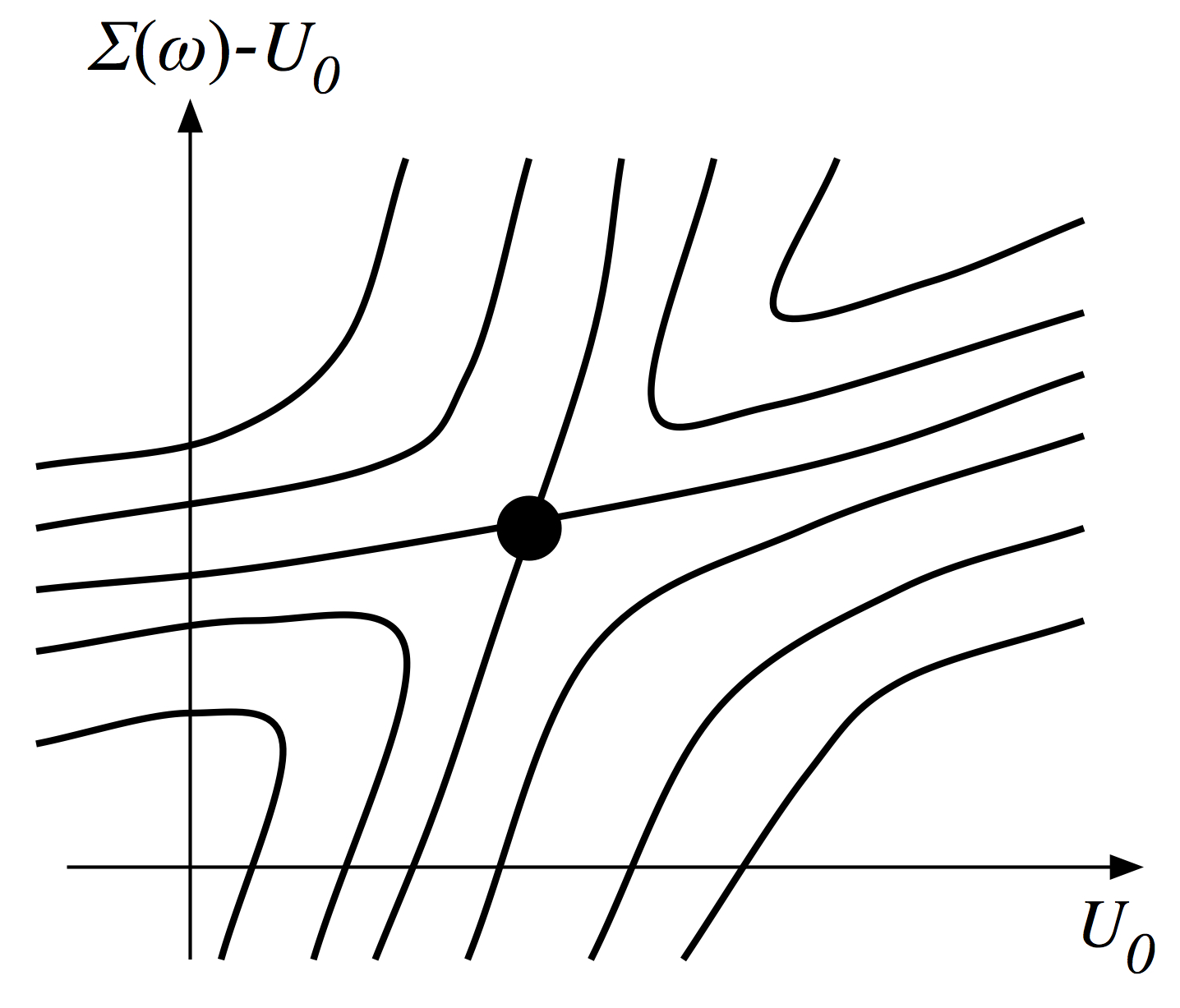}
\caption{Schematic figure showing the simplest likely scenario for the Klein total energy functional.  What is shown are level curves of the total energy functional $F[G]$ as a function of the self-energy $\Sigma_{xc}(\omega)$ that determines the Green's function $G$ via the Dyson Eq.~(\ref{eq:dysoneq}) (not necessarily self-consistently).  The horizontal axis represents self-energies that are static and Hermitian operators $U_0$ that will generate non-interacting Green's functions $G_0$ via Eq.~(\ref{eq:G0U0relation}).  The vertical axis represents the deviation of the self-energy from a static and Hermitian operator.  The black circle represents the extremum of the total energy functional corresponding to the true self-consistent self-energy which must occur for a dynamic and non-Hermitian $\Sigma_{xc}(\omega)$ since $F$ has no extremum along the horizontal axis.}
\label{fig:extremumschematic}
\end{figure}

Alas, our proof above shows that choosing the proper form of nonlocality is not straightforward.  The na\"ive idea of extremizing the total energy functional over all non-local $U_0$ to find the ``best'' $G_0$ leads to unphysical results.  For example, the simple and appealing idea of holding the wave functions $\psi_n(x)$ fixed and varying the eigenenergies $\epsilon_n$ to find an improved band structure via optimizing the total energy leads to the unphysical minimum. Of course, we know that there is an exact $G$ that extremizes the total energy functional and solves the Dyson equation self-consistently: there are examples of such calculations in the literature for model systems such as atoms, small molecules, or the electron gas \cite{GWheg1,GWheg2,GWheg3,GWheg4,DahlenLeeuwenvonBarth06}.  The problem is that the energy functional has no extremum in the subspace of Green's functions that correspond to non-interacting Green's functions.   To put this in pictures, the simplest likely scenario is illustrated schematically in Figure~\ref{fig:extremumschematic} where one assumes that the total energy functional has a single extremum corresponding to the true Green's function.  Obviously, that extremum must occur for a dynamic and non-local self-energy $\Sigma_{xc}(x,x',\omega)$.

The situation here is quite different from DFT where one can represent the electron density in terms of occupied non-interacting wave functions and where an unconstrained minimization over allowed densities, or equivalently over any set of occupied orthonormal non-interacting wave functions, leads to a single minimum with the correct ground-state energy \cite{HK,KS}. For the Klein functional, the analogous approach fails completely because an unconstrained optimization over non-interacting Green's functions drives the system to an unphysical minimum with negative infinite energy.  The analogous situation in DFT would be the (fictitious) situation where the correct ground-state density locates a minimum but the total energy functional has no minimum or lower bound when evaluated on a subset of allowed densities.

In our mind, there are two basic ways to overcome this hurdle. One could decide to solve the Dyson equation for the true self-consistent $G$ and avoid the instabilities.  However, this requires storing the entire Green's function $G(x,x',\omega)$ --- a full matrix as a function of continuous frequency $\omega$ --- which is highly prohibitive in terms of storage and computation and unlikely to lead to a method that will deal with realistic systems with many atoms in the near future.  The other approach is to stick with the simple and appealing picture of a noninteracting $G_0$ generated by a static and non-local $U_0$ but to constrain the extremization so as to avoid unphysical behavior.   In other words, one constrains the allowed forms of $U_0$ in some way.  In this light, the use of strictly local $V_{xc}$, the QS$GW$, and the scCOHSEX approaches can be viewed as schemes where one manually imposes physical constraints on the potential $U_0$ in order to avoid pathologies.  As noted above, when constraining $U_0$ to be a local potential, the available evidence strongly argues that the Klein functional will have a minimum \cite{DahlenvonBarthPRB04, DahlenLeeuwenvonBarth06, HellgrenvonBarth07}.  However, the open question is how to improve beyond these choices and to find better constraints that still avoid pathologies. This will involve coming up with some type of presently unknown metric to rank the various types of constraints.

\section{Approximate $\Phi_{xc}^{GW}$ based on the screened interaction}
\label{sec:approxPhic}

Although the rewriting represented by Eq.~(\ref{wprewrite}) is exact, there are a number of reasons to look for other expressions.  Ideally, it would be nice to rewrite $\Phi_{xc}^{GW}$ in terms of the central quantity in $GW$ which is the screened Coulomb interaction $W$.  One reason is to remove the dependence on the plasmon description and instead to deal only with interband transitions and screening.  Another reason is that dielectric functions and screening are relatively well understood and studied objects for which a variety of approximations and computational approaches exist.  As we show in Section~\ref{sec:cohsex} when deriving the COHSEX and related energy expressions, working with the screened interaction allows one to naturally build in different types of physical insights.  Before wading into the derivation, we highlight the endpoint:  they key result is Eq.~(\ref{Wtrewrite}).

We begin with Eq.~(\ref{Phixcepsilon}) and re-expand the logarithm,
\[
\Phi_{xc}^{GW}[G_0] = 
E_X[\rho_0]  - \frac{1}{2}
\sum_{j=2}^\infty \frac{1}{j}\int_{-\infty}^\infty \frac{d\omega}{2\pi i}\, tr \Big\{ [VP(\omega)]^j \Big\} \,.
\]
We concentrate on the correlation part $\Phi_c=\Phi_{xc}-E_X$ and write each $P(\omega)$ as a sum over transition energies as per Eq.~(\ref{eq:PDeltadef}):
\[
\Phi_c^{GW} = - \frac{1}{2}
\sum_{j=2}^\infty \frac{1}{j}\int_{-\infty}^\infty \frac{d\omega}{2\pi i}\, tr \Big\{ \sum_{\Delta_1}V\Pi_{\Delta_1}(\omega)\sum_{\Delta_2}V\Pi_{\Delta_2}(\omega)\cdots\sum_{\Delta_j}V\Pi_{\Delta_j}(\omega)\Big\}\,.
\]
We define $I_j$ as the $j^{th}$ term in this series,
\begin{equation}
I_j  \equiv  -\frac{1}{2j}\int_{-\infty}^\infty \frac{d\omega}{2\pi i}\, tr \Big\{ \sum_{\Delta_1}V\Pi_{\Delta_1}(\omega)\sum_{\Delta_2}V\Pi_{\Delta_2}(\omega)\cdots\sum_{\Delta_j}V\Pi_{\Delta_j}(\omega)\Big\}\,.
\label{Ij}
\end{equation}
We close the contour integral over the lower complex $\omega$ half-plane (the final result is unchanged if we choose the upper half-plane).  Since each factor $\Pi_\Delta(\omega)$ is given by
\[
\Pi_\Delta(\omega) = \frac{2\Delta}{(\omega-\Delta)(\omega+\Delta)}\sum_\delta \ket{\delta}\bra{\delta}\,,
\]
the integrand for $I_j$ is analytic everywhere in the lower half-plane except when $\omega=\Delta$ for some transition energy $\Delta$.  Since $\Pi_\Delta(\omega)$ diverges as $(\omega-\Delta)^{-1}$ close to such a point, the integrand for $I_j$ has divergences of type $(\omega-\Delta)^{-k}$ where $1 \le k \le j$ and $k$ labels how many of the $\{\Delta_1,\ldots,\Delta_j\}$ happen to be coincident.  This leads us to write $I_j$ as a sum over sets of contributions $I_j^k$ labeled by the divergence index $k$,
\[
I_j = \sum_{k=1}^j I_j^k
\]
where
\begin{equation}
I_j^k \!= \! -\frac{1}{2j}\!\oint\! \frac{d\omega}{2\pi i} \sum_\Delta 
{\sum_{q_0,\ldots,q_k}}^\prime
tr \Big\{
[VP_\Delta(\omega)]^{q_0} V\Pi_\Delta(\omega)[VP_\Delta(\omega)]^{q_1} \cdots V\Pi_\Delta(\omega)[VP_\Delta(\omega)]^{q_k}
\Big\}\,.
\label{Ijk}
\end{equation}
The contour integral is along the real axis and closed over the lower complex $\omega$ half-plane, $P_\Delta(\omega)$ is the polarization with transition energy $\Delta$ missing,
\[
P_\Delta(\omega)= P(\omega) - \Pi_\Delta(\omega) = \sum_{\Delta'\ne \Delta}\Pi_{\Delta'}(\omega)\,,
\]
and $\{q_0,\cdots,q_k\}$ are any nonnegative integers restricted to  sum to $j-k$,
\[
\sum_{l=0}^k q_l = j-k\,.
\]
The prime over the second sum denotes this restriction on $\{q_0,\cdots,q_k\}$.  
Eq.~(\ref{Ijk}) is to be understood as follows: to get a contribution to $I_j^k$, we must have $k$ of the $j$ transition energies in Eq.~(\ref{Ij}) have the same value which we call $\Delta$ while all the remaining $j-k$ transition energies must be different from $\Delta$; most generally, the first $q_0$ factors in Eq.~(\ref{Ij}) have transitions differing from $\Delta$, then there is a transition at $\Delta$, then $q_1$ follow which differ, followed by another at $\Delta$, \etc; summing over all $\Delta$ and all possible $\{q_l\}$ includes all the possibilities.

Using the cyclicity of the trace, we combine the first $q_0$ and last $q_k$ terms,
\[
I_j^k = 
-\frac{1}{2j}\oint \frac{d\omega}{2\pi i} \sum_\Delta 
{\sum_{q_0,\ldots,q_k}}^\prime
tr \Big\{ V\Pi_\Delta(\omega)[VP_\Delta(\omega)]^{q_1}V\Pi_\Delta(\omega) \cdots V\Pi_\Delta(\omega)[VP_\Delta(\omega)]^{q_0+q_k}
\Big\}\,,
\]
and then replace the $\Pi_\Delta$ by explicit sums over transitions to get
\beq
I_j^k & = &  -\frac{1}{2j}\oint \frac{d\omega}{2\pi i} \sum_\Delta 
\frac{(2\Delta)^k}{(\omega^2-\Delta^2)^k}
{\sum_{q_0,\ldots,q_k}}^\prime\sum_{\delta_1,\cdots,\delta_k}\nonumber\\
& & \qquad \qquad \qquad \qquad
tr \Big\{ V\ket{\delta_1}\bra{\delta_1}[VP_\Delta(\omega)]^{q_1}V\ket{\delta_2}\bra{\delta_2} \cdots V\ket{\delta_k}\bra{\delta_k}[VP_\Delta(\omega)]^{q_0+q_k}
\Big\}\\
& = &  -\frac{1}{2j}\oint \frac{d\omega}{2\pi i} \sum_\Delta 
\frac{(2\Delta)^k}{(\omega^2-\Delta^2)^k}
{\sum_{q_0,\ldots,q_k}}^\prime\sum_{\delta_1,\cdots,\delta_k}
\bra{\delta_1}[VP_\Delta(\omega)]^{q_1}V\ket{\delta_2}\cdots \bra{\delta_k}[VP_\Delta(\omega)]^{q_0+q_k}V\ket{\delta_1}\,,
\eeq
where the transition $\{\delta_1,\cdots,\delta_k\}$ sum only over those with energy $\Delta$.

In the above expression for $I_j^k$, there is no separate dependence on $q_0$ or $q_k$ but only on their sum $r=q_0+q_k$.  There are $r+1$ possibilities for $q_0$ and $q_k$ at fixed $r$.  Summing over them generates a multiplicative factor of $r+1$.  We rename $r$ back to $q_k$ and have
\begin{multline*}
I_j^k = -\frac{1}{2j}\oint \frac{d\omega}{2\pi i} \sum_\Delta 
\frac{(2\Delta)^k}{(\omega^2-\Delta^2)^k}
{\sum_{q_1,\ldots,q_k}}^\prime\sum_{\delta_1,\cdots,\delta_k}\\
\bra{\delta_1}[VP_\Delta(\omega)]^{q_1}V\ket{\delta_2}
\bra{\delta_2}[VP_\Delta(\omega)]^{q_1}V\ket{\delta_3}
\cdots
 (q_k+1)\bra{\delta_k}[VP_\Delta(\omega)]^{q_k}V\ket{\delta_1}\,.
\end{multline*}
We sum over one fewer $\{q_l\}$, namely from $q_1$ to $q_k$, where the prime indicates that $q_1+\cdots+q_k=j-k$ as before.

The $q_l$ are not treated symmetrically in the above expression:  $q_k$ is singled out by the extra factor $q_k+1$ because we chose to eliminate $q_0$.  However, the final result for $I_j^k$ has the same value if instead we single out another $q_l$ to have the $q_l+1$ factor because we can rearrange the $k$ multiplied factors and permute the $q_l$ via relabeling.  Therefore, we symmetrize by summing over all $k$ choices of $q_l$ (being singled out) and dividing by $k$.  This amounts to averaging the $q_l+1$ factor:
\[
\frac{1}{k}\sum_{l=1}^k (q_l+1) = \frac{1}{k}(j-k+k) = \frac{j}{k}
\]
which happily cancels the $1/j$ factor.  So we now have the symmetric expression
\[
I_j^k = 
-\frac{1}{2}\oint \frac{d\omega}{2\pi i} \sum_\Delta 
\frac{(2\Delta)^k}{k\cdot(\omega^2-\Delta^2)^k}
{\sum_{q_1,\ldots,q_k}}^\prime
\sum_{\delta_1,\cdots,\delta_k}\\
\bra{\delta_1}[VP_\Delta(\omega)]^{q_1}V\ket{\delta_2}
\cdots\bra{\delta_k}[VP_\Delta(\omega)]^{q_k}V\ket{\delta_1}\,,
\]
We are now ready to sum over $j$ to eliminate the restriction over the $\{q_l\}$, \ie\ remove the prime.  To do this, we reorder the $j$ and $k$ sums in the correlation energy
\[
\Phi^{GW}_{c} = \sum_{j=2}^\infty \sum_{k=1}^j I^k_j =  
\sum_{k=1}^\infty \sum_{j=k}^\infty I^k_j - I^1_1
\]
where we added and subtracted  $I^1_1=\frac{1}{2} \sum_t\bra{t}V\ket{t}$.  The inner sum over $j$ removes the constraint over the $q_l$ so 
\[
\sum_{j=k}^\infty I_j^k = 
-\frac{1}{2}\oint \frac{d\omega}{2\pi i} \sum_\Delta 
\frac{(2\Delta)^k}{k\cdot(\omega^2-\Delta^2)^k}
{\sum_{q_1,\ldots,q_k}}\sum_{\delta_1,\cdots,\delta_k}
\bra{\delta_1}[VP_\Delta]^{q_1}V\ket{\delta_2}\cdots\bra{\delta_k}[VP_\Delta]^{q_k}V\ket{\delta_1}\,.
\]
Each factor of $[VP_\Delta)]^{q_l}V$ can be summed separately to yield an identical result, and in each case we are summing the geometric series for $(1-x)^{-1}$:
\begin{equation}
W_\Delta(\omega) \equiv \sum_{q_l=0}^\infty [VP_\Delta(\omega)]^{q_l}V = (I-VP_\Delta(\omega))^{-1}V  \,.
\end{equation}
We have defined the modified screened interaction $W_\Delta(\omega)$ for which the interband transition energy $\Delta$ is missing from the screening action.  Thus the sum over $j$ has yielded
\[
\sum_{j=k}^\infty I_j^k = 
-\frac{1}{2}\oint \frac{d\omega}{2\pi i}\sum_\Delta \frac{(2\Delta)^k}{k\cdot(\omega^2-\Delta^2)^k}
\sum_{\delta_1,\cdots,\delta_k}\bra{\delta_1}W_\Delta(\omega)\ket{\delta_2}\cdots\bra{\delta_k}W_\Delta(\omega)\ket{\delta_1}\,.
\]
We define a square matrix $\overline{W_\Delta(\omega)}$ that contains the matrix elements of $W_\Delta(\omega)$ among all the degenerate transitions $\{\delta_1,\cdots,\delta_k\}$ of energy $\Delta$:
\[
\overline{W_\Delta(\omega)}_{p,q} \equiv \bra{\delta_p}W_\Delta(\omega)\ket{\delta_q}\,,
\]
which allows us to compactify the above expression as
\begin{equation}
\sum_{j=k}^\infty I_j^k = 
-\frac{1}{2}\oint \frac{d\omega}{2\pi i}\sum_\Delta \frac{(2\Delta)^k}{k\cdot(\omega^2-\Delta^2)^k}
tr\Big\{\left(\overline{W_\Delta(\omega)}\right)^k\Big\}\,.
\label{eq:exactcontourintegral}
\end{equation}

The result of Eq.~(\ref{eq:exactcontourintegral}) is exact and summing it over $k$ will recover the same answer as the exact result in Eq.~(\ref{wprewrite}) albeit expressed in a completely different manner.  We can, in principle, perform the contour integral.  On the lower complex $\omega$ half-plane, the integrand has a $k^{th}$ order pole at $\omega=\Delta$ as well as a large number of other poles coming from $W_\Delta(\omega)$ which physically correspond to the screening modes or plasma frequencies.  We would have to sum over all residues to obtain the integral exactly.   However, the resulting expression is unwieldy and can not be simplified in any meaningful manner known to us.

Therefore, we make a physical approximation to make the integral tractable.  The basic approximation is in the spirit of the COHSEX approximation:  we assume that the physically important plasma frequencies are at much higher energies than the dominant interband transition energies.  This is generally reasonable for solids and extended systems where plasmons are strongly collective modes with high frequencies due to the long range of the Coulomb interaction.  In other words, we assume that the screening dynamics is much faster than the interband dynamics.  We return to this point at the end of Section~\ref{sec:cohsex} where we discuss what type of physics is and is not included in this type of approximation.
  Separately, we provide some evidence of the relatively good quality of this type of approximation for atomic systems in Section~\ref{sec:atoms}.  

Mathematically, this approximation means that we ignore the contributions of the residues coming from $W_\Delta(\omega)$ itself and instead include only the residue of the low-energy pole at $\Delta$.  If $\tilde\omega$ is the energy of a typical pole of $W_\Delta(\omega)$, then the neglected terms are proportional to powers of the dimensionless ratio $\Delta/\tilde\omega$.  Therefore, if we write the integral as a power series in $\Delta/\tilde\omega$, our approximation amounts to keep only the leading terms of order $(\Delta/\tilde\omega)^0$.  Equivalently, we pretend that $W_\Delta(\omega)$ is smooth and analytic so that the only poles in the integral of Eq.~(\ref{eq:exactcontourintegral}) come from the denominator at $\omega=\Delta$.  

Since $W_\Delta(\omega)$ is built from the polarization $P_\Delta(\omega)$, both are missing transitions at energy $\Delta$ and are well behaved at and about $\omega=\Delta$.  We use Eq.~(\ref{cauchy})  to find
\[
\sum_{j=k}^\infty I^k_j  \approx \frac{1}{2}
\sum_\Delta \frac{(2\Delta)^k}{ k!}
\frac{d^{k-1}}{d\omega^{k-1}}\left[
\frac{1}{(\omega+\Delta)^k}\cdot tr\Big\{\left(\overline{W_\Delta(\omega)}\right)^k\Big\}\right]
\Big|_{\omega=\Delta}\,.
\]
Again, due to the assumption of the smoothness of $W_\Delta(\omega)$ at low frequencies, its derivatives at $\omega=\Delta$ are also assumed negligible compared to the derivatives of $(\omega+\Delta)^{-k}$: this amounts to discarding terms with positive powers of $\Delta/\tilde\omega$.  So we arrive at
\[
\sum_{j=k}^\infty I^k_j  \approx \frac{1}{2}
\sum_\Delta \frac{(2\Delta)^k}{k!}\cdot tr\Big\{\left(\overline{W_\Delta(\Delta)}\right)^k\Big\}\cdot
\frac{d^{k-1}}{d\omega^{k-1}}\left[
\frac{1}{(\omega+\Delta)^k}\right]
\Big|_{\omega=\Delta}\,.
\]
Taking the $k-1$ derivatives and evaluating at $\omega=\Delta$ yields
\beq
\sum_{j=k}^\infty I^k_j  & \approx & \frac{1}{2}
\sum_\Delta \frac{(2\Delta)^k}{k!}\cdot
tr\Big\{\left(\overline{W_\Delta(\Delta)}\right)^k\Big\}
\cdot\frac{(-1)^{k-1}(2k-2)!}{(2\Delta)^{2k-1}(k-1)!}\\
& \approx & \frac{1}{2}\sum_\Delta \Delta\cdot
tr\left\{\left(-\frac{\overline{W_\Delta(\Delta)}}{2\Delta}\right)^k\cdot
\frac{(2k)!}{(k!)^2\cdot(1-2k)}\right\}\,.
\eeq
Interestingly, we recognize this as a term in the Taylor series for $\sqrt{1+x}$,
\[
\sqrt{1+x} = \sum_{n=0}^\infty \left(-\frac{x}{4}\right)^n\cdot\frac{(2n)!}{(n!)^2(1-2n)}\,.
\]
We sum our approximate expression over all $k$ to get
\beq
\sum_{k=1}^\infty \sum_{j=k}^\infty I^k_j  & \approx & 
\frac{1}{2}\sum_\Delta \Delta\cdot tr\left\{
\sqrt{I+\frac{2\overline{W_\Delta(\Delta)}}{\Delta}}
-I\right\} \\
& \approx &  \frac{1}{2}\sum_\Delta tr \left\{ \sqrt{I \Delta^2+2\Delta\overline{W_\Delta(\Delta)}} - I\Delta \right\}\,.
\eeq
where a matrix square root are understood.  Putting this together with $I^1_1$, we have our main result for the approximate rewriting of the correlation energy:
\begin{equation}
\Phi^{GW}_{c} \approx \frac{1}{2}\sum_\Delta
tr\Big\{ \left[I\Delta^2+2\Delta\overline{W_\Delta(\Delta)}\right]^{1/2}\Big\}-\frac{1}{2}\sum_t \Big(\Delta_t + \bra{t}V\ket{t}\Big)\,.
\label{Wtrewrite}
\end{equation}
We have achieved our objective of summing the entire series for the $GW$-RPA correlation energy approximately and writing it as the sum of expectations of a  screened interaction over interband transitions.  The modified screened interaction $W_\Delta$ is built from a modified polarizability $P_\Delta$ which includes polarization contributions from of all interband electron-hole fluctuations except for those at energy $\Delta$.  This means the expression is self-interaction corrected in that excited electron-hole pairs at energy $\Delta$ do not screen themselves but are only screened by the other transitions.

Another perspective on this self-interaction correction is provided by identifying contributions to Eq.~(\ref{Wtrewrite}) that are exact and do not depend on the high-frequency screening approximation.  These contributions are the terms with $j=k$ ($I_j^j$ terms) because these terms have no $VP_\Delta$ factors since $q_0=\cdots=q_k=0$ is forced.  Therefore, only the bare Coulomb interaction is relevant for them and they represent a weak screening limit.  If we include only these terms, the high-frequency approximation is unnecessary since all $W_\Delta(\omega)$ are replaced by $V$, and Eq.~(\ref{Wtrewrite}) becomes
\[
\Phi^{GW}_{c} \rightarrow \frac{1}{2}\sum_\Delta
tr\Big\{ \left[I\Delta^2+2\Delta\overline{V}\right]^{1/2}\Big\}-\frac{1}{2}\sum_t \Big(\Delta_t + \bra{t}V\ket{t}\Big)\,.
\]
The matrices under the square roots are precisely the block-diagonals of the RPA/Casida $\Omega^2$ matrix of Eq.~(\ref{eq:Omega2matrix}) with  degenerate transition energies.  Namely, this formula is the expression we would obtain if we had taken the exact expression of Eq.~(\ref{wprewrite}) and computed $\omega_p$ by only solving the block diagonal parts of the RPA/Casida Eq.~(\ref{RPACasida}).  This provides a different viewpoint on why the screened interaction $W_\Delta$ appearing in Eq.~(\ref{Wtrewrite}) must be based on a polarizability that does not include any transitions at energy $\Delta$:  interactions of transitions with energy $\Delta$ among themselves are already included exactly by this formula with no screening (\ie\ when $W_\Delta=V$).  So the screened interaction in Eq.~(\ref{Wtrewrite}) must be responsible for capturing the couplings among transitions of differing energies which is precisely why transitions at $\Delta$ do not contribute to $W_\Delta$.

At this point, a great deal of simplification is achieved if the only degeneracies present are normal ones (\ie\ those due to a symmetry such as spin degeneracy for an spin unpolarized system or molecular or crystalline symmetry in real space).  Generically, we expect this to be the case as any weak perturbation will remove an accidental degeneracy.  For the case of normal degeneracy, the degenerate subspace spanned by the transitions $\{\delta_p\}$ with energy $\Delta$ transforms as an irreducible representation of the symmetry group.  On the other hand, the screening matrix $W_\Delta(\omega)$ must transform as the identical representation (\ie, as a scalar).  Thus the only non-zero entries in the matrix $\overline{W_\Delta(\omega)}$ will be the diagonal entries, and the diagonals will all be equal by symmetry.  The matrix square root is trivial in this diagonal basis and Eq.~(\ref{Wtrewrite}) becomes
\begin{equation}
\Phi^{GW}_{c} \approx \frac{1}{2}\sum_t
\sqrt{\Delta_t^2+2\Delta_t\bra{t}W_{\Delta_t}(\Delta_t)\ket{t}}-\Delta_t
 -\bra{t}V\ket{t}\,.
\label{Wtrewritediag}
\end{equation}
This relation can be viewed as a scalar version of the more general matrix Eq.~(\ref{Wtrewrite}).
In the unusual case of accidental degeneracies, the matrix $\overline{W_\Delta(\omega)}$ will have off-diagonal entries and it is necessary to evaluate the matrix square root of Eq.~(\ref{Wtrewrite}).  Since these degeneracies are rare, the matrices in question will be small and using any method (such as diagonalization) will not impact the computational load.

We note that these approximate forms of the $GW$-RPA correlation energy also suffer from the unboundedness problem discussed in Section~\ref{sec:unbounded} that plagues the exact $GW$-RPA correlation energy of Eq.~(\ref{wprewrite}).   For example, consider scaling all interband transitions by a factor $\lambda$, so $\Delta_t\rightarrow\lambda\Delta_t$, and then sending $\lambda$ to zero.  As long as matrix elements of $W_\Delta$ remain finite,  the correlation energies of Eqs.~(\ref{Wtrewrite},\ref{Wtrewritediag}) both diverge to negative infinity for the same reason that Eq.~(\ref{wprewrite}) diverges.  (In fact, when all interband transition energies are scaled to zero, the system will show very effective metallic screening and the matrix elements of $W_\Delta$ will actually go to zero.)  We present numerical evidence of this divergence in Section~\ref{sec:atoms}.

We end this section with some comments about the actual usage of Eqs.~(\ref{Wtrewrite},\ref{Wtrewritediag}).  Computationally, using these expressions requires re-computation of the screening for each transition energy.  In a na\"ive implementation,  the full matrix is recomputed, and since each computation of $P$ scales as $O(N^4)$, the overall computational load for evaluating Eq.~(\ref{Wtrewrite}) is $O(N^6)$ and thus no better than the exact diagonalization approach of Eqs.~(\ref{wprewrite}) and (\ref{RPACasida}).  Furthermore, the sums in Eqs.~(\ref{Wtrewrite},\ref{Wtrewritediag}) are very difficult to converge numerically because transition energies in the continuum are very closely spaced and thus $W_\Delta(\omega)$ will have contributions from many poles near $\omega=\Delta$.  Converging such a sum requires a very dense representation of the continuum together with some regularization procedure to control the very large contributions from nearby poles.  In Section~\ref{sec:atoms}, we discuss this problem for atomic systems.   Thus, the utility of Eqs.~(\ref{Wtrewrite},\ref{Wtrewritediag}) is not for numerical computation per se but rather for deriving new approximations, as shown in the next section.

\section{COHSEX-type correlation functionals}
\label{sec:cohsex}

In addition to being closed form expressions for the correlation energy, Eqs.~(\ref{Wtrewrite}) and (\ref{Wtrewritediag}) are based on the screened interaction.  As explained above, literal implementation of these formulae is computationally expensive and numerically difficult to converge.  Their utility, rather, is in analytical work where one can contemplate a variety of approximations to the screening that incorporate different physical effects.  As an example, we present the simplest approximation that leads to COHSEX and its associated $\Phi_c$.  During the process, we will derive a ladder of related approximations.  Looking ahead, the relative quality of a number of the approximations and simplifications are tested numerically on atoms in Section~\ref{sec:atoms}.

We base our derivation on the scalar expression of Eq.~(\ref{Wtrewritediag}) for simplicity of presentation.  The full matrix expression of Eq.~(\ref{Wtrewrite}) can also be used but produces more complex-looking results with the same physical content.
In what follows, we use $W_t(\omega)$ as a shorthand for $W_{\Delta_t}(\omega)$.  

The first approximation is to work within the basic idea of COHSEX:  replace the dynamic screening matrix by a static one at $\omega=0$.  This leads to a first approximate form,
\begin{equation}
\Phi^{stat}_{c}= \frac{1}{2}\!\sum_t\!\!
\sqrt{\Delta_t^2+2\Delta_t\bra{t}W_t(0)\ket{t}}-\Delta_t
 -\bra{t}V\ket{t}\,.
\label{Phistat}
\vspace*{-2ex}
\end{equation}
There is a simple relation between matrix elements of the usual screened interaction $W(\omega)=\varepsilon^{-1}(\omega)V$ and the $W_t(\omega)$ appearing here that we exploit to compute the static screening.  The following matrix identity 
\[
u^\dag(A-xuu^\dag)^{-1}u = u^\dag A^{-1}u/(1-u^\dag A^{-1}ux)
\]
for vector $u$, matrix $A$, and scalar $x$, allows to make the connection.  With $A=V^{-1}-P_{\Delta_t}$, $u=\ket{t}$ and $x=2\Delta_t/(\omega^2-\Delta_t^2)$, we find
\begin{equation}
\bra{t} W_t(\omega)\ket{t} =\frac{\bra{t}W(\omega)\ket{t}}{1+\bra{t}W(\omega)\ket{t}\cdot2\Delta_t(\omega^2-\Delta_t^2)^{-1}}\,.
\label{WtWrelation}
\end{equation}
For static screening, we have
\begin{equation}
\bra{t} W_t(0)\ket{t} =\frac{\bra{t}W(0)\ket{t}}{1-2\bra{t}W(0)\ket{t}/\Delta_t}\,.
\label{Wt0W0relation}
\end{equation}
Therefore, in terms of computational complexity, evaluation of Eq.~(\ref{Phistat}) requires one calculation of the matrix $W(0)$, computation of its diagonal elements in the $\ket{t}$ basis, and use of Eq.~(\ref{Wt0W0relation}).  For a system of $N$ atoms, this entire project scales as $O(N^4)$.  For comparison, the original expression of Eq.~(\ref{Wtrewrite}) requires evaluation of $W_t(\Delta_t)$ at each interband energy $\Delta_t$ and scales as $O(N^6)$ in a na\"ive implementation.

The next approximation expands the square root in Eq.~(\ref{Phistat}) in powers of 
$\bra{t}W_t(0)\ket{t}/\Delta_t$ and keeps the lowest order term,
\begin{equation}
\Phi^{stat}_c =\frac{1}{2}\sum_t
\bra{t}W_t(0)-V\ket{t} + O\Big(\langle W \rangle^2/\Delta\Big)\,.
\vspace*{-2ex}
\label{eq:PhistatWt0sqrtexpand}
\end{equation}
This approximation requires that $|\bra{t}W_t(0)\ket{t}/\Delta_t|\ll1$.  Section~\ref{sec:atoms} provides numerical results for atoms where we show that this expansion of the square root is actually quite accurate.    Since screening is rather weak in the localized atomic limit, we expect this approximation to work even better in extended solids which provide much more effective screening of the Coulomb interaction.

The third approximation is to replace the matrix element $\bra{t}W_t(0)\ket{t}$ by $\bra{t}W(0)\ket{t}$.  This also creates an error of $O(\langle W \rangle^2/\Delta)$ and should be applicable in the same cases as the previous approximation.  The result is
\begin{equation}
\Phi^{stat}_c =\frac{1}{2}\sum_t
\bra{t}W(0)-V\ket{t} + O\Big(\langle W \rangle^2/\Delta\Big)\,.
\vspace*{-2ex}
\label{eq:PhistatW0sqrtexpand}
\end{equation}
Completeness $\sum_c \ket{\psi_c}\bra{\psi_c}= I - \sum_v \ket{\psi_v}\bra{\psi_v}$ permits us to rewrite $\Phi_c^{stat}$ using only valence states.  Putting $E_X$ back and dropping terms of order $O(\langle W \rangle^2/\Delta)$, 
\begin{equation}
\Phi_{xc}^{stat} = \frac{1}{2}\int  dx\, \rho_0(x,x)W^{pol}(x,x,0)
-\frac{1}{2} \int dx\int  dx'\, |\rho_0(x,x')|^2W(x,x',0) 
\label{COHSEXenergy}
\end{equation}
where $W^{pol}=W-V$ is the induced (polarization) part of $W$.  The first term is the Coulomb hole energy (COH) and the second term the screened exchange energy (SEX).   Differentiating versus the non-interacting density matrix $\rho_0$  yields the static self-energy $\Sigma_{xc}^{stat} = \delta \Phi_{xc}^{stat}/\delta\rho_0$,
\begin{equation}
\Sigma_{xc}^{stat}(x,x') =  \frac{1}{2}W^{pol}(x,x,0)\delta(x-x') \\ -\rho_0(x,x')W(x,x',0) + O(\delta W/\delta \rho_0)\,.
\label{COHSEXsigma}
\end{equation}
This is the COHSEX self-energy if we drop the derivative term $\delta W/\delta \rho_0$.  Ignoring the derivatives underlies the successful BSE approach for optical excitations \cite{BSE} and excited states forces \cite{BSEforces}: the successes of the BSE approach suggests that these terms are not significant in practice.  In brief, the COHSEX self-energy of Eq.~(\ref{COHSEXsigma}) is approximately associated with and derived from the exchange-correlation energy of Eq.~(\ref{COHSEXenergy}).  The key approximations underlying COHSEX are the static screening limit {\it and} the assumption that the matrix element $\bra{t}W(0)\ket{t}$ for a transition $t$ is much smaller than its energy $\Delta_t$.  Given $W$, Eqs.~(\ref{COHSEXenergy}) and (\ref{COHSEXsigma}) depend only on the valence states through $\rho_0$.  Having the energy functional available in Eq.~(\ref{COHSEXenergy}), we also know what missing terms we must add to the COHSEX self-energy of Eq.~(\ref{COHSEXsigma}) in order to make it a variational scheme.

To improve COHSEX, we step backwards through the approximations.  The first stop is to use Eq.~(\ref{Phistat}):  this is still a static approximation but doesn't assume that $\langle W\rangle/\Delta$ is small.  To go beyond Eq.~(\ref{Phistat}), we need dynamic screening.  In order to avoid the $O(N^6)$ scaling of the direct implementation of Eq.~(\ref{Wtrewrite}), we can contemplate the following scheme:  we  relate $W_t(\Delta_t)$ by to $W(\Delta_t)$ via Eq.~(\ref{WtWrelation}) and then approximate $W$ by using sum rules as per plasmon-pole models to approximate the frequency dependence of $W(\omega)$ \cite{HL,plasmonpoleLH,plasmonpoleEF,plasmonpoleZTCLH}.  Alternatively, we can use model dielectric screening functions \cite{modelepsLL,modelepsHL,modelepsCDSRB} to construct approximate $W(\omega)$ and then use Eq.~(\ref{WtWrelation}) to find $W_t$.  Regardless of the specifics,  having a total energy expression allows us to find the associated self-energy through differentiation.

We conclude this section with some observations on what is and is not included when using the approximate expressions derived in this and the previous section.  We expect these approximations to certainly include static and localized Coulombic effects  automatically.  This includes screening effects of the medium as well as localized Coulombic physics:  even the simplest COHSEX self-energy of Eq.~(\ref{COHSEXsigma}) and COHSEX correlation energy of Eq.~(\ref{COHSEXenergy}) contain the non-local density matrix $\rho_0$ that projects the action of the screened Coulomb interaction $W$ onto the occupied states. If the occupied states are derived from localized states such as $3d$ or $4f$ orbitals of transition metals, then the projection is automatically onto this manifold.  Namely, already at the COHSEX level, we expect to recover the benefits of an LDA+U type treatment.  This is no surprise since a static and localized approximation to the $GW$ self-energy yields the kernel of the LDA+U approach \cite{LDApU}.  As an added benefit, a COHSEX type approach should automatically include a properly screened $U$ parameter, as opposed to $U$ being an externally chosen parameter in the usual LDA+U method. The dynamic formulae of Eqs.~(\ref{Wtrewrite}) and (\ref{Wtrewritediag}) extend these results to include the frequency dependence of the screening.  What is missing from these approximate results are the contributions to the RPA correlation energy which are physically distinct from those stemming from screened interband transitions:  these are the neglected residue contributions in Eq.~(\ref{eq:exactcontourintegral}) from the poles from the screening modes themselves.  Unfortunately, at present we are unable to provide simple physical examples of situations where these neglected contributions play the dominant role and our key approximation to fail for basic physical reasons.  This question is a subject of present investigation.

\section{Numerical tests: atoms}
\label{sec:atoms}

In this section, we describe numerical results on atomic systems.  The aim of this section is not to present an exhaustive and comprehensive treatment.  That is the subject of a future investigation.  Rather, the main aim is to demonstrate the numerical efficacy of the plasmon formula Eq.~(\ref{RPACasida}) for the $GW$-RPA correlation energy and to test the main approximation of high frequency screening used in the approximate results of Eqs.~(\ref{Wtrewrite}) and (\ref{Wtrewritediag}).

Our atomic code uses a standard non-relativistic approach.  The atomic eigenfunctions are assumed to take a spherical form:  for an eigenstate with spin index $\sigma=\pm1$, the spatial part is $R_{nl\sigma}(r)Y_{lm}(\theta,\phi)$.  The radial part $R_{nl}(r)$ is represented numerically on a radial grid of exponentially-spaced points \cite{expradialgrid,NIST}.  The spherical harmonics $Y_{lm}$ are included up to a some maximum angular momentum $l_{max}$, typically $l_{max}=3$ below.  When building up the one-particle density matrix $\rho_\sigma$ for spin $\sigma$,
\[
\rho_{\sigma}(r,\theta,\phi,r',\theta',\phi')  = \sum_{n,l,m} f_{nlm\sigma} R_{nl\sigma}(r)R_{nl\sigma}(r')Y_{lm}(\theta,\phi)Y_{lm}(\theta',\phi')^*\,,
\]
we allow for $m$-dependent state fillings $f_{nlm\sigma}\in\{0,1\}$.
The local spin-density approximation (LSDA) \cite{LSDA} or unrestricted Hartree-Fock (HF) \cite{HFbook} equations are solved by minimizing the appropriate total energy over arbitrary occupied orthonormal radial functions.  We have tested our code with available high quality LSDA \cite{NIST} and HF \cite{accurateHF} data to ensure agreement to at least one part in 10$^8$ in total energies.  When computing various correlation energies below, we make a spherical approximation which is consistent with assuming spherical eigenfunctions.  This means that when we compute the polarizability ($P$ or $P_t$) and the screened interaction ($W$ or $W_t$), we assume the fillings have no $m$-dependence (\ie, shell occupancies $f_{nl\sigma}$ instead of orbital occupancies $f_{nlm\sigma}$):  different $m$-channels will not interact so that taking both $l$ and $m$ as good quantum numbers is a self-consistent assumption.  However, when using matrix elements of $W$ or $W_t$ to compute a contribution to the energy such as in Eq.~(\ref{Wtrewrite}), we do include the full $m$-dependence of the occupancies $f_{nlm\sigma}$.

\begin{figure}[t]
\includegraphics[width=5in]{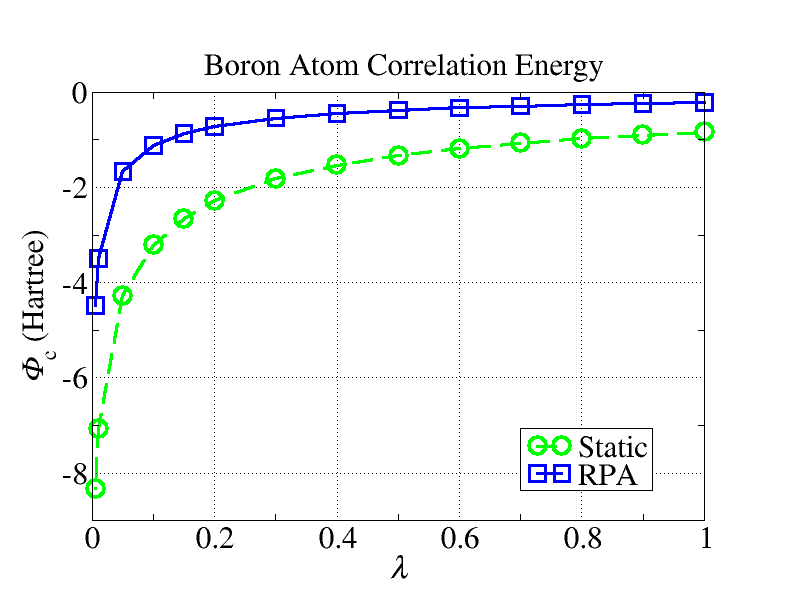}
\caption{Correlation energy $\Phi_c$ as a function of the scaling $\lambda$ of the transition energies for the boron atom.  Single-particle energies are from a ground-state LSDA calculation.  The blue squares with solid line are the exact RPA correlation energies from Eq.~(\ref{wprewrite}) when using scaled transition energies in the RPA/Casida Eq.~(\ref{RPACasida}).  The green squares with dashed line are correlation energies from the static approximation of Eq.~(\ref{Phistat}) using the scaled transition energies for computing the screening.  The dashed or solid lines are guides for the eye.  Clearly, the energy decreases monotonically with decreasing $\lambda$ to large and unphysical negative values with no extrema along this path.
}
\label{fig:Eclambda}
\end{figure}

We begin by showing direct numerical evidence for the unboundedness of the $GW$-RPA correlation energy $\Phi_c^{GW}$ that was proven in Section~\ref{sec:unbounded}.  We consider the case of the boron atom with configuration $1s^22s^22p^1$.  The occupied single-particle states and energies are found self-consistently within LSDA for a radial grid extending to $r_{max}=10$ Bohr radii.  After finding the self-consistent potential for the LSDA ground state, we generate the lowest 300 states for each angular momentum and spin channel.  We then compute and tabulate the Coulomb matrix elements in the RPA/Casida Eq.~(\ref{RPACasida}) within this basis of transitions and find the correlation energy of Eq.~(\ref{wprewrite}) through direct diagonalization.  Starting with this information, we then scale all transition energies $\Delta_t$ uniformly by a factor $\lambda$ where $0<\lambda<1$ so that $\Delta_t\rightarrow\lambda\Delta_t$.  We use the scaled transition energies in the RPA/Casida Eq.~(\ref{RPACasida}) to find the corresponding plasma frequencies and then use Eq.~(\ref{wprewrite}) to compute $\Phi_c^{GW}$ as a function of $\lambda$.  Figure~\ref{fig:Eclambda} shows the correlation energy as a function of $\lambda$.  As is evident, the correlation energy becomes very negative and unphysically large in magnitude.  Furthermore, the curve is monotonic in $\lambda$:  starting with the reasonable $\lambda=1$ energy, one can drive the correlation energy to arbitrarily negative values along a continuous path with no extremum along this coordinate.  As discussed in Section~\ref{sec:approxPhic}, the approximate forms also have this unphysical behavior, and the Figure shows the example of the static approximation of Eq.~(\ref{Phistat}).

We now move onto more pragmatic questions, the first being the convergence properties of the plasmon form of Eq.~(\ref{wprewrite}).  As discussed above, the plasmon form for the $GW$-RPA correlation energy is more physically transparent than the standard integral expression of Eq.~(\ref{Phixcepsilon}).  In addition, it tuns out that it converges much more rapidly.  This is demonstrated for the cases of the helium atom ($1s^2$ configuration) and boron atom ($1s^22s^22p^1$ configuration) in Table~\ref{table:HeBconvg}.   To generate the data in this table, an atomic LSDA calculation is run to self-consistency for each atom for a radial grid of $r_{max}=7$ Bohr radii for He and $r_{max}=10$ for B, and the lowest 300 eigenstates and eigenenergies of the Kohn-Sham Hamiltonian are computed and tabulated.  The $N_{max}$ lowest-energy eigenstates are then used to evaluate the correlation energies, and convergence is monitored by increasing $N_{max}$.

For the plasmon form, the RPA/Casida Eq.~(\ref{RPACasida}) is solved within the basis of transitions generated by the $N_{max}$ single-particle states via direct diagonalization.  For the integral form, the computations are more involved.  First, for numerical stability, the integral along the real axis is changed to along the imaginary $\omega$ axis (renamed $\beta$) via a Wick rotation.  Using the identity $tr \ln A = \ln \det A$ for a matrix $A$, the integral is then 
\begin{equation}
\Phi^{GW}_{c} = - \frac{1}{2} \sum_t \bra{t}V\ket{t}
+\frac{1}{2}\int_{-\infty}^\infty \frac{d\beta}{2\pi}\, \ln \det  \varepsilon(i\beta)\,.
\label{eq:Ecbeta}
\end{equation}
The matrix determinant is computed within the subspace spanned by the single-particle states.  Second, the upper limit is reduced to a finite value $\beta_{max}$ and the integral is discretized with spacing $\Delta\beta$.  As we desire the integral in the limits $\Delta\beta\rightarrow0$ and $\beta_{max}\rightarrow\infty$, two extrapolations are performed.  For fixed $\beta_{max}$, the integral is evaluated for a series of $\Delta\beta$, and Richardson extrapolation is performed to $\Delta\beta=0$.  Next, these extrapolated values are themselves Richardson extrapolated to $\beta_{max}=\infty$ by noting that for large $\beta$, $\varepsilon(i\beta)=I+A\beta^{-2}+O(\beta^{-4})$ so that the neglected integral from $\beta_{max}$ to $\infty$ is proportional to $\beta_{max}^{-1}$ to leading order.  These extrapolation procedures are straightforward and unproblematic because the integrand is smooth as a function of $\beta$.  The doubly extrapolated values are listed in Table~\ref{table:HeBconvg}.  

Clearly, the plasmon form has superior convergence properties when compared to the integral form, and, additionally, does not require any particular extrapolation or discretization procedures.  Helium presents a rather simple case:  more complex atoms (or molecules) will require progressively larger basis sets to converge the integrals and  the plasmon form should prove even more useful in practice.  A hint of this is provided by comparing He to B in Table~\ref{table:HeBconvg}.  For B, the integral form of the correlation energy converges more slowly, and the calculations at $N_{max}=300$ were already quite demanding in terms of time (and patience).  The plasmon form, again, shows rapid convergence versus $N_{max}$.

\begin{table}
\caption{Comparison of convergence of $GW$-RPA correlation energies for the He and B atoms.  The RPA correlation energy is computed using the standard integral of Eq.~(\ref{eq:Ecbeta}) (third column) or the plasmon form of Eq.~(\ref{wprewrite}) (fourth column).  All energies are in Hartrees.  $N_{max}$ is the number of single-particle state included in the calculations for each angular momentum and spin channel.  The single particles energies and wave functions are from the LSDA.}\label{table:HeBconvg}
\begin{tabular}{cccc}
\hline\hline
Atom\ \ \  & $N_{max}$ & \ \ \ integral\ \ \  & \ \ \ plasmon\ \ \ \\
\hline
He & 25 &  -0.1038 & -0.0804\\
He & 50  & -0.0916 & -0.0805\\
He & 100  & -0.0849 & -0.0806\\
He & 150  & -0.0830 & -0.0806\\
He & 200  & -0.0819 & -0.0806\\
\hline
B & 50 & -0.4126 & -0.2129\\
B & 100 & -0.3600 & -0.2171\\
B & 200 & -0.2844 & -0.2175\\
B & 300  & -0.2584 & -0.2175\\
\hline\hline
\end{tabular}
\end{table}

We now examine the accuracy of the approximate forms derived in Sections \ref{sec:approxPhic} and \ref{sec:cohsex} above.  In Tables~\ref{table:HeLSDAvsHF} and \ref{table:BLSDAvsHF} we present these correlation energies for the helium and boron atoms using LSDA or HF wave functions and eigenenergies.  These are to be compared to the exact RPA energy of Eq.~(\ref{wprewrite}) in the second to last row in each table.  For completeness, we also include the other energy terms to show their relative importance and their variation with single-particle theory, discussed further below. The Hartree energy reported in the table is the one based on the actual, non-spherical electron density (\ie\ based on $m$-dependent occupancies $f_{nlm\sigma}$ as opposed to the spherical density in most DFT atomic calculations).
\begin{table}
\caption{Various components of the total energy of the helium atom.  Results are reported for single-particle wave functions and eigenenergies coming from self-consistent LSDA or Hartree-Fock (HF) calculations.  Energies are in Hartree. Except for the (*) values, all calculations use a radial grid of size $r_{max}=18$ Bohr radii and $N_{max}=257$ single-particle states for each angular momentum channel.
The starred (*) values are difficult to converge, and the values in the table are uncertain to within $\pm$0.001:  see text for details.  For reference, the exact (CI) non-relativistic correlation energy is also shown as the last entry.
}
\label{table:HeLSDAvsHF}
\begin{tabular}{lcccc}
\hline\hline
Energy component & Equation &  \ \ \ \ LSDA \ \ \ \  & \ \ \ \  HF  \ \ \ \ &  Difference\\
\hline
Kinetic & & 2.768 & 2.862 & 1.0\%\\
Electron-ion & & -6.626 & -6.749 & 1.0\%\\
Hartree & & 1.996 & 2.052 & 1.0\%\\
Fock exchange & (\ref{eq:EX}) & -0.998 & -1.026 & 1.0\%\\
$\Phi_c$: $W(0)$ instead of $W_t(0)$ & (\ref{eq:PhistatW0sqrtexpand}) & -0.318 & -0.255 & 25\%\\
$\Phi_c$: square root approx. & (\ref{eq:PhistatWt0sqrtexpand}) & -0.311 & -0.248 & 25\%\\
$\Phi_c$: static approx. & (\ref{Phistat}) & -0.313 & -0.250 & 25\%\\
$\Phi_c$: main approx. & (\ref{Wtrewritediag}) & -0.060* & -0.048* & 25\% \\
$\Phi_c$: RPA & (\ref{wprewrite}) & -0.081 & -0.064 & 27\%\\
$\Phi_c$: CI (exact) Ref.~\cite{CI1} & & \multicolumn{2}{c}{-0.0420}  \\
\hline\hline
\end{tabular}
\end{table}

\begin{table}
\caption{Various components of the total energy of the boron atom.  Results are reported for single-particle eigenfunctions and eigenenergies coming from self-consistent LSDA or HF calculations.  All energies are in Hartree. Except for the (*) values, all calculations use a radial grid of size $r_{max}=40$ Bohr radii and $N_{max}=400$ single-particle states for each angular momentum channel.
The starred (*) values are difficult to converge, and the values in the table are actually uncertain to within $\pm$0.01:  see text for details.  For reference, the exact (CI) non-relativistic correlation energy is also shown as the last entry.
}
\label{table:BLSDAvsHF}
\begin{tabular}{lcccc}
\hline\hline
Energy component & Equation & \ \ \ \ LSDA \ \ \ \  & \ \ \ \  HF \ \ \ \  & Difference\\
\hline
Kinetic & & 24.173 & 24.530 & 1.5\%\\
Electron-ion & & -56.520 & -56.900 & 0.67\%\\
Hartree & & 11.534 & 11.590 & 0.49\%\\
Fock exchange & (\ref{eq:EX}) & -3.712 & -3.749 & 1.0\%\\
$\Phi_c$: $W(0)$ instead of $W_t(0)$ & (\ref{eq:PhistatW0sqrtexpand}) & -0.869 & -0.717 & 21\%\\
$\Phi_c$: square root approx. & (\ref{eq:PhistatWt0sqrtexpand}) & -0.820 & -0.680 & 21\%\\
$\Phi_c$: static approx. & (\ref{Phistat}) & -0.835 & -0.692 & 20\%\\
$\Phi_c$: main approx. & (\ref{Wtrewritediag}) & -0.30* & -0.11* & 270\% \\
$\Phi_c$: RPA & (\ref{wprewrite}) & -0.217 & -0.171 & 27\%\\
$\Phi_c$: CI (exact) Ref.~\cite{CI2}  & & \multicolumn{2}{c}{-0.125}  \\
\hline\hline
\end{tabular}
\end{table}

We begin by considering our main approximate form of Eq.~(\ref{Wtrewritediag}).  As we can see from comparing to the exact RPA energy, and especially when comparing to the static versions, the basic approximation underlying Eqs.~(\ref{Wtrewrite}) and (\ref{Wtrewritediag}) is a relatively good one:  the absolute correlation energy of Eq.~(\ref{Wtrewritediag}) differs by at most 0.1 Ha from the exact RPA one.  This shows that the fundamental approximation of assuming high-frequency screening is reasonable even in atoms.  For solids and extended systems where the Coulomb interaction shows true long-ranged behavior (as opposed to atoms), plasma modes are of higher energies than interband energies and the situation should be further improved.

As explained above, the approximate form of Eq.~(\ref{Wtrewritediag}) is as computationally expensive to calculate as the exact RPA plasmon form but is additionally very difficult to converge.  To obtain the values in the tables, we had to perform the following steps simultaneously:  (i) increase the size of the radial axis to make for a denser continuum, (ii) increase the number of one-particle states entering the calculation to ensure a fixed level of convergence with increasing radial axis, and (iii) exclude contributions to $W_t(\Delta_t)$ from transitions $t'$ that were within a small energy window $\delta$ of $\Delta_t$ (\ie, $|\Delta_{t'}-\Delta_t|<\delta$) while sending $\delta\rightarrow0$.

Moving on to the approximations that assume static screening, we see that they overestimate the importance of correlations and give too negative values uniformly.    We discuss the reason for this in the next paragraph, but in the meantime we see that once the static approximation is made, the various forms for the correlation energy are quite similar.  For example, the difference between the square root form of Eq.~(\ref{Phistat}) and its series expansion in Eq.~(\ref{eq:PhistatWt0sqrtexpand}) is small.  The smallness of the differences simply means that the ratio $\langle W(0)\rangle/\Delta$ is small:  \eg, for boron we find the largest value of the ratio $\langle W(0)\rangle/\Delta$ is 0.2 and is achieved for the $2s$-$2p$ transition. Given how far all these static correlation energies are from the dynamic answer of Eq.~(\ref{Wtrewritediag}) or the exact RPA answer of Eq.~(\ref{wprewrite}), one can take all these static approximations to be basically of equal accuracy.

\begin{figure}[t!]
\includegraphics[width=6.2in]{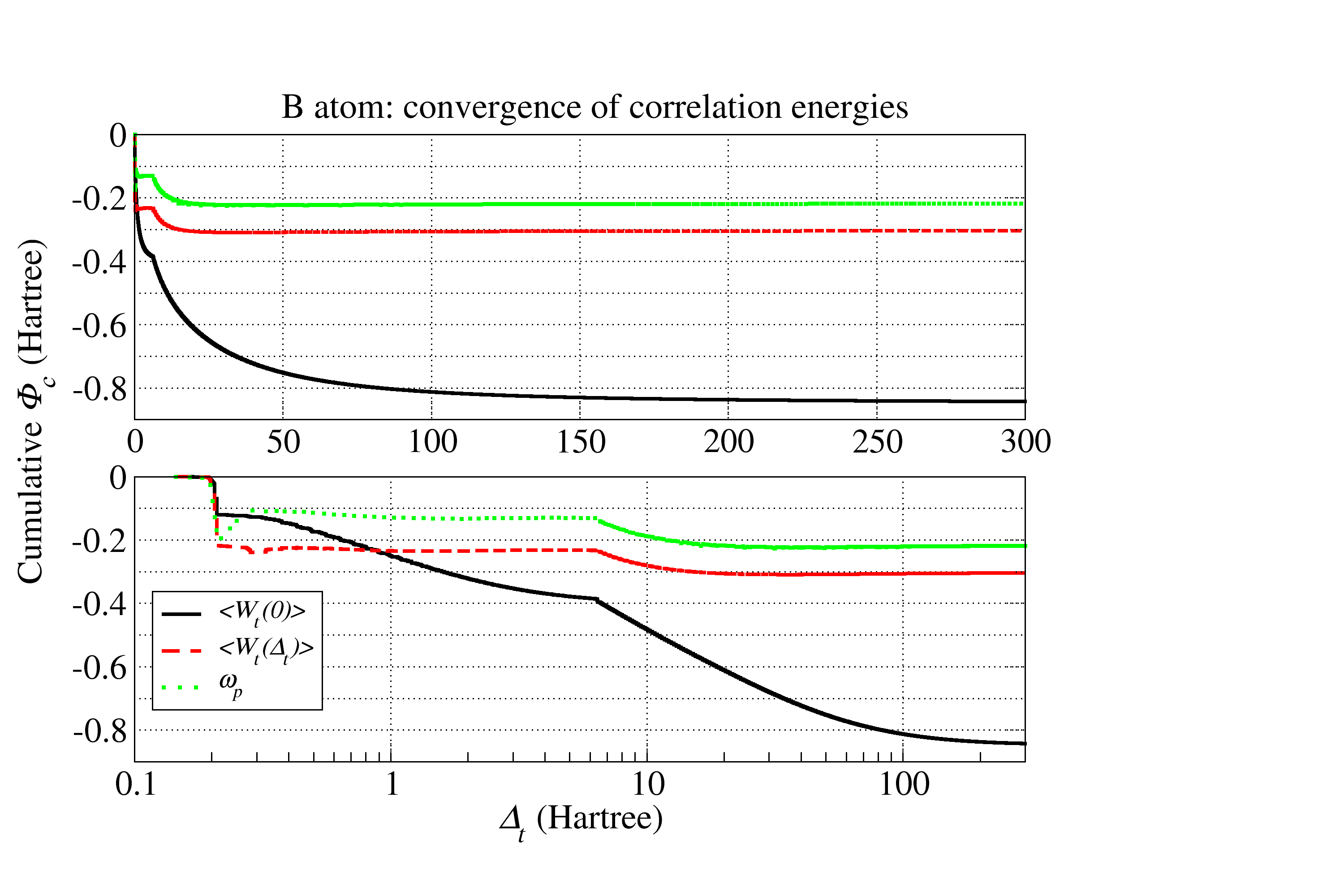}
\caption{Convergence of correlation energies for the boron atom. Both plots show the  cumulative sum of the contributions to the correlation energies (vertical) for transitions and/or plasma modes up to some given energy $\Delta_t$ (horizontal).  LSDA wave functions and eigenenergies are used with a radial grid of size $r_{max}=40$ Bohr radii and $N_{max}=400$ eigenstates.  The lowest solid black curve is for the static approximation of Eq.~(\ref{Phistat}), the middle red dashed curve is for the dynamic approximation of Eq.~(\ref{Wtrewritediag}), and the uppermost green dotted curve is for the exact RPA formula of Eq.~(\ref{wprewrite}).  The plots show the same data with the only difference being a linear (top plot) or a logarithmic scale (bottom plot) of the horizontal axis.  The dominant transitions in LSDA boron are the 2$s$--2$p$ at 0.21 Ha and 1$s$-2$p$ at 6.4 Ha, both visible in the lower plot as energies where the correlation contributions have a sudden jump.
}
\label{fig:convgcumsums}
\end{figure}

The reason the static approximations overestimate the magnitude of the correlation energy is easy to understand.  Figure~\ref{fig:convgcumsums} shows how three representative correlation energies converge for the case of atomic boron: the static formula of Eq.~(\ref{Phistat}), the dynamic formula of Eq.~(\ref{Wtrewritediag}), and the exact RPA formula of Eq.~(\ref{wprewrite}).  What is shown is the correlation energy contributions summed up to a given transition energy (or plasmon energy for the exact case).  In all cases, we see that the contributions to the correlations are small, then become large at a certain energy, and then become smaller again.  In atomic boron at LSDA level, there are two physically important transitions:  the dominant 2$s$--2$p$ at 0.21 Ha and then the weaker 1$s$--2$p$ at 6.4 Ha.  As the transition energy sweeps through each atomic transition, we see large contributions to the correlation energies at first, but then the transition is ``exhausted'' and the contributions become small again.  

The obvious difference between the three curves is that for the static case all the contributions are always negative, tend to be large in magnitude, and keep adding up to yield a large negative value.  On the other hand, for the dynamic approximation and the exact result, the initial low-energy contributions are large and negative, may then become slightly positive, but rapidly become small in magnitude.  To understand this difference, we write the modified screened interaction as $W_t(\omega) = \varepsilon_t(\omega)^{-1}V$ in terms of a modified dielectric function $\varepsilon_t(\omega)$.  Solving the RPA/Casida Eq.~(\ref{RPACasida}) with transition $t$ missing gives us a set modified plasma modes with energies $\tilde \omega_p$ and mode functions $\ket{\tilde p}$.  Based on Eq.~(\ref{epsinvedef}), we have the dynamic and static matrix elements 
\begin{eqnarray*}
\bra{t}W_t(\Delta_t)\ket{t} & = & \bra{t}V\ket{t} + \sum_p \frac{2\tilde\omega_p |\bra{t}V\ket{\tilde p}|^2}{\Delta_t^2-\tilde\omega_p^2}\,,\\
\bra{t}W_t(0)\ket{t} & = & \bra{t}V\ket{t} - \sum_p \frac{2\tilde\omega_p |\bra{t}V\ket{\tilde p}|^2}{\tilde\omega_p^2}\,.
\end{eqnarray*}
Clearly, the static formula always gives negative correlation contributions that become small only when the matrix elements $\bra{t}V\ket{\tilde p}$ become small.  On the other hand, the dynamic case has a denominator that changes sign from positive to negative as $\Delta_t$ sweeps through the plasma energies, and the denominator itself gets large when $\Delta_t$ gets large.  The dynamic screening behavior of the plasmons is missing in the static formula which assumes that no matter what the transition energy, the plasmons can screen it adiabatically.  This is obviously erroneous for transitions where $\Delta_t\gtrsim\tilde\omega_p$.  Therefore, the static formula can be improved by adding some dynamic behavior in the screening even if done approximately.  This is an example of what was meant in Section~\ref{sec:cohsex} regarding the use of plasmon-pole models or model dielectric functions to improve the static COHSEX.

We end this section with some observations on the results in Tables~\ref{table:HeLSDAvsHF} and \ref{table:BLSDAvsHF} and their dependence on the single-particle theory.  The parts of the total energy that depend only on the single-particle orbitals, \ie\ the kinetic, electron-ion, Hartree, and Fock exchange energies, depend weakly on the choice of LSDA versus HF single-particle orbitals, changing at most $\sim$1\%.  This is not surprising {\it a posteriori} as visual comparison of the radial functions show small differences.  However, the correlation energies depend more strongly on the choice of single-particle theory, and this is due to the relatively large differences between the single-particle energies.  The HF-based correlation energies are smaller than the LSDA simply because HF transition energies are larger than LSDA:  \eg, for atomic boron the important $2s$-$2p$ transition is at 0.210 Ha in LSDA but at 0.232 Ha in HF; larger transition energies mean weaker screening and thus weaker correlation.  As per Section~\ref{sec:unbounded}, it matters greatly whether the Green's function $G_0$ is generated by a local or non-local potential.

The true many-body Green's function $G$ obeys Dyson's Eq.~({\ref{eq:dysoneq}) and is thus generated by a non-local (and dynamic) self-energy $\Sigma_{xc}(x,x',\omega)$.  {\it A priori}, we would expect that a static but non-local potential $U_0(x,x')$ should generate a decent non-interacting $G_0$ which should be close to the true $G$, certainly closer than one generated by a static local potential.  However, as discussed in Section~\ref{sec:unbounded}, choosing the ``best'' $U_0$ via an optimization of the total energy functional is problematic unless constraints are imposed on $U_0$. The outstanding theoretical problem is what constraints to impose and which ones are the ``best'' ones.  This obviously involves creating and justifying  metrics that tells us how good each choice of constraints will be in practice.

\section{Summary and Outlook}
\label{sec:conclusions}

Our work has focused on the correlation energy functional with Luttinger-Ward theory (specifically the Klein functional) within the $GW$-RPA approximation.  The Green's functions used in the energy functional are of non-interacting form.  The main findings in this work are threefold.  First, we present the exact rewriting of the $GW$-RPA correlation energy functional in Eq.~(\ref{wprewrite}) in terms of differences between plasma and interband energies.  This form is directly amenable to computation, shows good convergence properties in the atomic tests, and has prospects for having its computational scaling improved by use of matrix square root algorithms.  Second, we describe the approximate rewriting of the $GW$-RPA correlation energy functional in Eqs.~(\ref{Wtrewrite}) and (\ref{Wtrewritediag}) where the correlation energy is written as a sum of screened interband transition contributions;  the main approximation is to assume that the dominant screening dynamics are much faster than the key interband dynamics; atomic tests show that the approximation is good numerically.  These approximate forms then lead to a ladder of approximations where the COHSEX is the simplest one possible.  Third, we show, analytically and with numerical examples for atoms, that if one restricts the Green's function to be of non-interacting form generated by Hermitian non-local potentials, the $GW$-RPA correlation energy has no lower bound over this set of Green's functions; nor does it have an extremum.

Going forward, the exact rewriting and its good convergence properties --- coupled with algorithmic development --- should pave the wave for wider application of the $GW$-RPA correlation functional to materials systems.  The multitude of approximate forms we present here will hopefully broaden and improve the types of approximate self-energies used in self-consistent band structure methods that go beyond the usual LDA or GGA treatments.  Simultaneously, having explicit energy functionals on hand means one can construct the self-energies in a variational manner:  the solution of the self-consistent equation for a particular self-energy will optimize the total energy functional from which it was derived.  However, the result on the unboundedness of the $GW$-RPA correlation energy over the space of all non-interacting Green's functions means that applying these Green's function approaches is not yet a rote exercise in applying standard unconstrained optimization algorithms.  Rather, the choice of non-local potential that generates the non-interacting Green's function must be constrained in some manner so as to avoid the pathological negative infinite correlation energy and to produce extrema.  The theoretical question is then to understand which constraints succeed and also to devise metrics to compare their quality and accuracy.

\begin{center}
{\bf Acknowledgements}
\end{center}

We wish to acknowledge helpful discussions or comments from Kris Delaney, Kieron Burke, Douglas A. Stone, John Tully and Mark van Schilfgaarde. This work has been supported primarily by the National Science Foundation under Grant No. MRSEC DMR 0520495.


\begin{thebibliography}{99}

\bibitem{HK} P. Hohenberg and W. Kohn, \emph{Phys. Rev.} \textbf{136}, B864 (1964).

\bibitem{KS} W. Kohn and L. Sham, \emph{Phys. Rev.} \textbf{140},  A1133 (1965).

\bibitem{LDA} J. P. Perdew and A. Zunger, \prb{23}, 5048 (1981).

\bibitem{GGA} J. P. Perdew, J. A. Chevary, S. H. Vosko, K. A. Jackson, M. R. Pederson, D. J. Singh and C. Fiolhais, \prb{46}, 6671 (1992).

\bibitem{LDAbadgaps} \emph{Theory of the Inhomogeneous Electron Gas},
  edited by S. Lundqvist and N. H. March (Plenum, New York 1983) and
  references therein.

\bibitem{HL} M. S. Hybertsen and S. G. Louie, \prb{34}, 5390 (1986).

\bibitem{LDApU} V. I. Anisimov, F. Aryasetiawan and A. I. Lichtenstein, {\it J. Phys. Cond. Matt.} {\bf 9}, 767 (1997).

\bibitem{DMFT1} A. Georges, G. Kotliar, W. Krauth and M. J. Rozenberg, \rmp{68}, 13 (1996).
 
\bibitem {DMFT2} G. Kotliar, S. Y. Savrasov, K. Haule, V. S. Oudovenko, O. Parcollet and C. A. Marianetti, \rmp{78}, 865 (2006). 

\bibitem{Hedin} L. Hedin, \prb{139}, A796 (1965). 

\bibitem{TDDFTvsGWBSE} G. Onida, L. Reining and A. Rubio, \rmp{74}, 601 (2002).

\bibitem{GWaeSiMnONiO} S. V. Faleev, M. van Schilfgaarde and T. Kotani, \prl{93}, 126406 (2004).

\bibitem{GWNiO} F. Aryasetiawan and O. Gunnarsson, \prl{74}, 3221 (1995).

\bibitem{TMOGWmodel} S. Massidda, A. Continenza, M. Posternak and A. Baldereschi, \prb{55}, 13494 (1997).

\bibitem{GWZnO} B. Kr\'alik, E. K. Chang and S. G. Louie, \prb{57}, 7027 (1998).

\bibitem{GWMgCaTiVO} A. Yamasaki and T. Fujiwara, \prb{66}, 245108 (2002).


\bibitem{GWNiO2} J.-L. Li, G.-M. Rignanese and S. G. Louie, \prb{71}, 193102 (2005).

\bibitem{QPSCGW} M. van Schilfgaarde, T. Kotani and S. Faleev, \prl{96}, 226402 (2006).

\bibitem{scgwCu2O} F. Bruneval et al., \prl{97}, 267601 (2006).

\bibitem{scCOHSEX} F. Bruneval, N. Vast and L. Reining, \prb{74}, 045102 (2006).

\bibitem{Furche} F. Furche, \emph{J. Chem. Phys.} {\bf 129}, 114105 (2008).

\bibitem{LW} J. M. Luttinger and J. C. Ward, {\it Phys. Rev.} {\bf 118}, 1417 (1960). 

\bibitem{GWheg1} U. von Barth and B. Holm, \prb{54}, 8411 (1996).

\bibitem{GWheg2} B. Holm and F. Aryasetiawan, \prb{56}, 12825 (1997).

\bibitem{GWheg3} B. Holm and U. von Barth, \prb{57}, 2108 (1998).

\bibitem{GWheg4} B. Holm, \prl{83}, 788 (1999).

\bibitem{GWhubbard} A. Schindlmayr, T. J. Pollehn and R. W. Godby, \prb{58}, 12684 (1998).

\bibitem{GWmolecules} F. Furche, \prb{64}, 195120 (2001).

\bibitem{GWH2hubbard} F. Aryasetiawan, T. Miyake and K. Terakura,  \prl{88}, 166401 (2002).

\bibitem{Miyake02} T. Miyake, F. Aryasetiawan, T. Kotani, M. van Schilfgaarde, M. Usuda and K. Terakura, \prb{66} 245103 (2002).

\bibitem{GWH2Be2} M. Fuchs and X. Gonze, \prb{65}, 235109 (2002).

\bibitem{GWSiNa} T. Miyake, F. Aryasetiawan, T. Kotani, M. van Schilfgaarde, M. Usuda and K. Terakura, \prb{66}, 245103 (2002).

\bibitem{DahlenvonBarthPRB04} N. E. Dahlen and U. von Barth, \prb{69}, 195102 (2004).

\bibitem{DahlenvonBarthJCP04} N. E. Dahlen and U. von Barth, {\it J. Chem. Phys.} {\bf 120}, 5826 (2004).

\bibitem{DahlenLeeuwenvonBarth06} N. E. Dahlen, R. van Leeuwen and U. von Barth, \pra{73} 012511 (2006).

\bibitem{Marinivdw06} A. Marini, P. Garc\'ia-Gonz\'alez and A. Rubio, \prl{96} 136404 (2006).

\bibitem{HellgrenvonBarth07} M. Hellgren and U. von Barth, \prb{76}, 075107 (2007).

\bibitem{HarlKresse08} J. Harl and G. Kresse, \prb{77} 045136 (2008).

\bibitem{Toulouserangesep09} J. Toulouse, I. C. Gerber, G. Jansen, A. Savin and J. G. \'Angy\'an, \prl{102} 096404 (2009).


\bibitem{Casida} M. E. Casida, Recent Advances in Density-Functional Methods, D. P. Chong ed. (World ScientiÞc, Singapore, 1995); Recent Developments and Applications of Modern Density Functional Theory, J. M. Seminario ed. (Elsevier, Amsterdam, 1996). 

\bibitem{Pines} D. Pines, Ch. 3 of {\it Elementary Excitations in Solids} (Addison-Wesley, 1963).

\bibitem{RMP} M. C. Payne, M. P. Teter, D. C. Allan, T. A. Arias and
J. D. Joannopoulos, \rmp{64}, 1045 (1992).


\bibitem{TDDFT1} E. Runge and E. K. U. Gross, \prl{52}, 997 (1984).

\bibitem{TDDFT2} E. K. U. Gross and W. Kohn, \prl{55}, 2850 (1985).

\bibitem{TDDFT3} E. K. U. Gross, \prl{57}, 923 (1986).

\bibitem{BSE} S. Albrecht, L. Reining, R. Del Sole and G. Onida, \prl{80}, 4510 (1998); L. X. Benedict,  E. L. Shirley and R. B. Bohn,
\prl{80}, 4514 (1998); M. Rohlfing and S. G. Louie, \prl{81}, 2312 (1998).

\bibitem{ShamSchluter} L. J. Sham and M. Schl\"uter, \prl{51} 1888 (1983).

\bibitem{LSS}R. W. Godby, M. Schl\"uter and L. J. Sham, \prb{37}, 10159 (1988).

\bibitem{BSElong} M. Rohlfing and S. G. Louie, \prb{62}, 4927 (2000).

\bibitem{Klein} A. Klein,  \emph{Phys. Rev.} {\bf 121}, 950 (1961).

\bibitem{Gilbert} T. Gilbert, \prb{12} 2111 (1975).

\bibitem{Harriman} J. E. Harriman, \pra{24} 680 (1981).

\bibitem{Casidavxcoptimal} M. E. Casida, \pra{51} 2005 (1995).

\bibitem{QPSCGWneginf} T. Kotani, M. van Schilfgaarde, S. V. Faleev and A. Chantis, \emph{J. Phys.: Cond. Matt.} {\bf 19}, 365236 (2007).

\bibitem{BSEforces} S. Ismail-Beigi and S. G. Louie, \prl{90}, 076401 (2003).

\bibitem{plasmonpoleLH} W. von der Linden and P. Horsch, \prb{37}, 8351 (1988).

\bibitem{plasmonpoleEF} G. E. Engel and B. Farid, \prb{47}, 15931 (1993).

\bibitem{plasmonpoleZTCLH} S. B. Zhang, D. Tom\'anek, M. L. Cohen, S. G. Louie and M. S. Hybertsen, \prb{40} 3162 (1989).

\bibitem{modelepsLL} Z. H. Levine and S. G. Louie, \prb{25} 6310 (1982).

\bibitem{modelepsHL} M. S. Hybertsen and S. G. Louie, \prb{37} 2733 (1988).

\bibitem{modelepsCDSRB} G. Cappellini, R. Del Sole, L. Reining and F. Bechstedt, \prb{47} 9892 (1993).


\bibitem{LSDA} U. von Barth and L. Hedin,  {\it J. Phys. C: Solid State Phys.} {\bf 4}, 1629 (1972); O. Gunnarsson and B. I. Lundqvist, \prb{13}, 4274 (1976); S. H. Vosko, L. Wilk and M. Nusair, \emph{Canadian J.Ê of Physics} \textbf{58}, 1200 (1980); S. H.Ê Vosko and L.Ê Wilk, \prb{22}, 3812 (1980).

\bibitem{expradialgrid} J. P. Desclaux, {\it Comp. Phys. Comm.} {\bf 1}, 216 (1969). 

\bibitem{NIST} S. Kotochigova {\it et al.} (2003), Atomic Reference Data for Electronic Structure Calculations (version 1.3). [Online] Available: http://physics.nist.gov/DFTdata [2009, December 2]. National Institute of Standards and Technology, Gaithersburg, MD.; S. Kotochigova, Z. H. Levine, E. L. Shirley, M. D. Stiles and C. W. Clark,  \pra{55}, 191 (1997); erratum {\bf 56}, 5191 (1997).

\bibitem{HFbook} C. Froese Fischer, {\it The Hartree-Fock Method for Atoms} (Wiley, New York, 1977).

\bibitem{accurateHF} C. F. Bunge, J. A. Barrientos, A. V. Bunge and J. A. Cogordan, \pra{46}, 3691 (1992).

\bibitem{CI1} E. R. Davidson, S. A. Hagstrom, S. J. Chakravorty, V. M. Umar and C. F. Fischer, \pra{44} 7071 (1991).

\bibitem{CI2} S. J. Chakravorty, S. R. Gwaltney, E. R. Davidson, F. A. Parpia and C. F. Fischer, \pra{47} 3649 (1993).


\end{thebibliography}
\end{document}